\documentclass[11pt, oneside]{article}   	
\usepackage{geometry}                		
\usepackage{graphicx}				
\usepackage{natbib}
\usepackage{mnemonic}	   
\usepackage{amssymb}
\usepackage{color}
\usepackage{ulem}
\date{}							

\newcommand{\fracb}[2]{\left(\frac{#1}{#2}\right)}

\newcommand{\eqb}{\begin{equation}}
\newcommand{\eqe}{\end{equation}}

\definecolor{forestgreen}{rgb}{0.13, 0.55, 0.13}
\definecolor{green(html/cssgreen)}{rgb}{0.0, 0.5, 0.0}
\definecolor{orange}{rgb}{1.0, 0.5, 0.0}

\def \simless {\mathbin{\lower 3pt\hbox{$\rlap{\raise 5pt
              \hbox{$\char'074$}}\mathchar"7218$}}}
\def \simgreat {\mathbin{\lower 3pt\hbox{$\rlap{\raise 5pt
              \hbox{$\char'076$}}\mathchar"7218$}}}

\newcommand{\zh}{z_{_{\rm h}}}
\newcommand{\rhead}{r_{_{\rm h}}}
\newcommand{\rL}{r_{_{\rm L}}}
\newcommand{\rj}{r_{_{\rm j}}}
\newcommand{\rc}{r_{_{\rm c}}}
\newcommand{\ph}{p_{_{\rm h}}}
\newcommand{\pL}{p_{_{\rm L}}}
\newcommand{\pj}{p_{_{\rm j}}}
\newcommand{\pc}{p_{_{\rm c}}}
\newcommand{\vh}{v_{_{\rm h}}}

\newcommand{\bh}{\beta_{_{\rm h}}}
\newcommand{\gh}{\Gamma_{_{\rm h}}}
\newcommand{\uh}{u_{_{\rm h}}}
\newcommand{\uj}{u_{_{\rm j}}}
\newcommand{\Gj}{\Gamma_{_{\rm j}}}

\newcommand{\bj}{\beta_{_{\rm j}}}
\newcommand{\thj}{\theta_{_{\rm j}}}
\newcommand{\Lj}{L_{_{\rm j}}}
\newcommand{\rext}{\rho_{_{\rm ext}}}
\newcommand{\aL}{a_{_{\rm L}}}

\makeatletter
\newcommand{\do@openE}[1]{%
  \mbox{\fontsize{#1}\z@\cmuserif\symbol{"0190}}%
}
\newcommand{\openE}{\mathord{\mathchoice
  {\do@openE\tf@size}
  {\do@openE\tf@size}
  {\do@openE\sf@size}
  {\do@openE\ssf@size}
}}
\makeatother

\begin{document}

\title{The dynamics of a highly magnetized jet propagating inside a star}

\author{Omer Bromberg$^{1,4}$, Jonathan Granot$^2$,
Yuri Lyubarsky$^3$, Tsvi Piran$^4$}


\maketitle

\centerline{\footnotesize ${^1}${Department of Astrophysical Sciences, Princeton University, 4 Ivy Ln., Princeton NJ 
08544, USA}}
\centerline{\footnotesize${^2}${Department of Natural Sciences, The Open University of Israel, P.O.B 808,
Ra'anana 43537, Israel}}
\centerline{\footnotesize${^3}${Physics Department, Ben-Gurion University, P.O.B. 653, 
Beer-Sheva 84105, Israel}}
\centerline{\footnotesize ${^4}${Racah Institute of Physics, The Hebrew University, Jerusalem 91904, 
Israel}}

\begin{abstract}
The collapsar model explains the association of long duration gamma-Ray Bursts  (GRBs)  
with stellar collapse. It involves a relativistic jet that forms at the core of a collapsing 
massive star. The jet penetrates the stellar envelope and the prompt GRB emission is 
produced once the jet is well outside the star. Most current models for generation of 
relativistic jets involve Poynting flux dominated outflows.  We explore here the propagation 
of such a jet through a stellar envelope. The jet forms a bow shock around  it. 
Energy dissipation at the head of this shock supplies energy to a cocoon that surrounds 
the jet. This cocoon exerts pressure on the jet and collimates it. While this description 
resembles the propagation of a hydrodynamic jets there are significant qualitative 
differences. Two Strong shocks, the reverse shock that slows down the hydrodynamic 
jet and the collimation shock that collimates it, cannot  form within the  Poynting flux 
dominated jet. As a result this jet moves much faster and dissipates much less energy 
while it crosses the stellar envelope. We construct here a simple analytic model that 
explores, self consistently, 
the jet-cocoon interaction and dynamics. Using this  model we determine 
the properties of the jet, including its velocity, propagation time and shape.  
\end{abstract}


\section{introduction}\label{sec:introduction}

Gamma-ray bursts (GRBs) are short and intense bursts of low energy gamma rays.
They involve powerful relativistic jets. 
Long GRBs are associated with death of massive stars.
The collapsar model\footnote{Note that we use here a  general
definition of the collapsar model in which it involves any central engine that launches a jet
within a collapsing star. This is regardless of the specific nature of the central engine. }
\citep{1993ApJ...405..273W, 1999ApJ...524..262M}
combined these two facts. According to this model
a compact object is formed at the center of a star following its core
collapse.  {This compact object launches a jet  that }
drills a hole through the star and breaks out through the surface. The   observed gamma-rays are emitted due to some
internal dissipation process  far from the surface of the star.
Clearly the propagation phase of the jet within the star is an essential  
ingredient of the model and in recent years
a lot of numerical  
\citep[e.g.][]{2001ApJ...550..410M,2003ApJ...586..356Z,2007ApJ...665..569M,
2009ApJ...699.1261M,2013ApJ...777..162M}
and analytic 
\citep[e.g.][]{2003MNRAS.345..575M,2005ApJ...629..903L,2007ApJ...665..569M,
2011ApJ...740..100B}
efforts was 
devoted to explore the propagation of a hydrodynamics jets within  stars.

{Most current relativistic jet  models are based in one way or another on a central  engine that generate a collimated
Poynting flux dominated outflow. This is partially motivated due to analogy with AGNs. 
For AGNs, Poynting flux is the only available option. In GRBs a thermally driven jet (a fireball) 
is also thermodynamically
possible. Still   it is generally expected that those will be less powerful than the 
electromagnetic ones
\citep[see e.g.][]{2013ApJ...766...31K}.
It is therefore important to investigate the properties of a magnetic
jet that propagates in the star, and whether its typical properties
agrees with the observational constraints.

Collimated relativistic MHD jets in stellar environments
were studied extensively both numerically and analytically under the
approximation of an axi-symmetric, steady,
non-dissipative flow 
\citep[e.g.][]{2007MNRAS.380...51K,2009MNRAS.394.1182K,
2008MNRAS.388..551T,2008AIPC.1054...71T,2009ApJ...699.1789T,
2008ApJ...679..990Z,
2009ApJ...698.1570L,2010ApJ...725L.234L,2011PhRvE..83a6302L,
2010NewA...15..749T,
2012MNRAS.426..595K}.
In these studies, however, either the shape of the jet or the profile of the confining
medium was predetermined. In a case of a predetermined pressure
\citet{2009ApJ...698.1570L,2010ApJ...725L.234L,2011PhRvE..83a6302L}
have shown analytically how the properties of the jet
are determined by a given distribution of the confining pressure, assuming a steady state.

In realistic  outflows, however, the confining pressure is built
during the course of the jet propagation, and its profile depends on the details
of this propagation. Therefore the properties of the propagating jet
and the confining pressure should be determined self consistently.
Recently, \citet{2013ApJ...764..148L} 
conducted an analytic, self consistent
analysis for the propagation of a magnetic jet in a medium.
Their model involves some constraining assumptions on the cocoon that result
in a cylindrical jet, having a cross section radius of the order of the light cylinder
radius of the central object. In additions they assume that kink instability grows
in the jet on a time scale which is comparable to a few light crossing time of the
jet width. 
This have lead them to conclude that the magnetic
the jet will be disrupted by kink instability deep in the star, and transform to a 
hydrodynamic jet.
However, as we show here, the jet is expected to be much wider and
it is most likely that the kink instability has no time 
to develop in the jet before it breaks out of the star. Therefore, we assume here 
that the jet remains axi-symmetric.

We study,  the dynamics of a Poynting dominated jet that propagates inside the star.
The dissipation of energy at the jet's head leads to the formation
of a hot cocoon that surround the jet. The cocoon  applies pressure on the jet and collimates it.
We build a self consistent analytic model
that follows the time evolution of the jet and the cocoon.
We show that the pressure in the cocoon is {typically} large enough to collimate the jet
close to the source, so that different parts of the jet maintain strong
causal connection with the jet's axis. In this case the poloidal magnetic field
is comparable to the toroidal field in the proper frame of the jet. This leads
to a  smooth transition of the jet material from a free
expansion state, near the engine, to a collated state.
We also show that the typical width of the propagating jet is of the order of a few 10s
light cylinder radii, which imply that the  main body of the jet is stable
to kink modes and that the jet is likely to survive  crossing  the star.
We compare our results with recent
numerical simulations of a highly magnetized jet \citep{BT14} and
 show that there is a good agreement
between the numerical and the analytic results.
Finally, we discuss the differences between the propagation of a magnetic jet and
a hydrodynamic jet,  focusing on the breakout time of the jet from
the star. We show that the propagation velocity of magnetic jets is relativistic
in most parts of the star (unlike hydrodynamic jets which propagate  typically at
sub relativistic velocities).

The paper is structured as follow:
In section 2 we describe the overall picture and lay down our basic assumptions.
We calculate, In section 3,  the propagation velocity of the jet and show
that it depends only on the properties of the confining medium near the jet's 
head. We then proceed to describe the geometry of the jet and it's 
cocoon (section 4), followed by a discussion on the conditions at the base of the jet and 
the collimation of the jet (section 5).
In section 6  we discuss the stability of the jet and its survival inside the star.
Finally, in section 7 we obtain  engine minimal activity time for a jet breakout. 
This quantity is of
outmost importance as it might be related to the observed plateau in the long GRB duration 
distribution \citep{2012ApJ...749..110B}.

\section{The overall picture and the model assumptions}\label{sec:general}

We work within the scope of the standard collapsar picture of GRBs, a central
engine (accreting black hole, or a millisecond magnetar) is producing a jet
that pushes its way through the progenitor star.
We assume that the jet is dominated by Poynting flux. The
properties of the jet are determined by the total luminosity, $\Lj$,
by the light cylinder radius, defined as $\rL =c/\Omega_0$, where $\Omega_0$ 
is the angular velocity of the field lines, and by the initial magnetization
{\citep[Michel magnetization;][]{1969ApJ...158..727M}}, 
$\sigma_0$, defined as the ratio of the Poynting flux and the rest mass flux.
The magnetization, $\sigma_0$ 
is in fact the maximal Lorentz factor 
achievable by the jet. Observations place a lower limit of 
$\Gamma_{max}\simgreat100$ 
\citep[e.g.][]{1995astro.ph..7114P,2001ApJ...555..540L,2008ApJ...677...92G},
implying $\sigma_0\simgreat100$.

When the jet propagates in the stellar envelope a bow shock
forms ahead of it. This shock heats the ambient medium forming a pressurized cocoon 
around the jet. 
The cocoon applies
pressure on the jet and collimates it.
Describing the properties of such system is an involved problem,
since the cocoon is produced by the jet while the jet's properties are determined
by the pressure distribution in the cocoon. Therefore the properties
of the jet and the cocoon must be found self consistently.

The overall morphology of the jet-cocoon system is sketched in 
Figure \ref{fig:jet}. 
We use cylindrical coordinates $(z,r)$, where the jet points along the $z$ axis, 
and the central source is at the origin. The jet's head is located at
$z=\zh$.  Coordinates along the jet axis are characterized either by 
the distance from the source, $z$, or by the distance from the head,
$\bar{z}=\zh-z$. The jet's width is denoted by $\rj(z)$ and the cocoon's
width by $\rc(z)$. The jet's material generally moves relativistically with a Lorentz factor 
$\Gj$. Therefore one can conveniently  describe the flow in two frames: the lab frame 
and the local comoving frame. Quantities measured in the local comoving frame 
are denoted by $'$. {When we discuss the properties of the jet's head
and the cocoon region near the head, we use a third reference frame, the rest 
frame of the head. We mark it by $''$ to distinguish it from the comoving
frame of the let material.}

Our model contains four regions: the
jet, the jet's head (shown in a zoom in on the left side of
Fig.~\ref{fig:jet}), the cocoon and the external medium. Parameters
relating to each  of these regions are marked with the indices j -
for the jet, h - for the head, c - for the cocoon, and `ext' - for the
external medium. The subscript ${\rm L}$ denotes
parameters on the jet's light cylinder.

When the plasma is injected at the base of the jet,
its internal pressure, $\pj$, is so large that initially it expands freely 
until the collimation point where the jet's pressure equals the cocoon's pressure, 
$\pj=\pc$. Above this point the jet is collimated by the 
 the cocoon's  pressure. 
  The transition to a collimated state is
accompanied by oscillations in the jet's radius around
the equilibrium state  \citep{2009ApJ...698.1570L}. These oscillations gradually decay in an expanding jet.
We ignore these oscillations here  and 
discuss only the average structure. The properties of the jet that is collimated by a given  external  pressure 
can be described as follows.

In a Poynting dominated jet the magnetic field beyond the light cylinder is dominated by 
the toroidal component. However when calculating the 
response of the jet to the pressure excreted
by the cocoon we must include the pressure contribution from the poloidal  components
as well.
The reason for that is that in a magnetic field that is predominantly azimuthal (toroidal),
the hoop stress is nearly counterbalanced by the electric force
\citep[see e.g.][]{2009ApJ...698.1570L,2011PhRvE..83a6302L}.
The outflow is governed by the residual small stress
$B_{\phi}^2-E^2\approx B'^2_{\phi}\sim {(B_{\phi}/\Gj)}^2$, therefore generally the pressure of the poloidal
 field cannot be   neglected.

The overall structure and the dynamics of the flow  depend crucially
on how well different parts of the jet can communicate
with each other across the flow 
\citep{2008ApJ...679..990Z,2009MNRAS.394.1182K,2009ApJ...698.1570L,2011PhRvE..83a6302L,
2009ApJ...699.1789T,2010NewA...15..749T,2011MNRAS.411.1323G}.
In our case the jet is so narrow that it remains strongly causally connected. This means that  the flow can communicate with the
boundary in a time that is shorter than the time it takes it to double
its radius. This implies the condition 
\begin{equation}
  \label{eq:ujthj}
  \uj\thj<1, 
\end{equation}
 where $\uj \equiv \Gj \bj$ is the spatial component of the four velocity and 
$\thj$ is  the opening angle of the flow.  In
this regime, the jet  maintains transverse magnetic
equilibrium at any distance from the source.
Here
the difference between the magnetic hoop stress and the electric force is counterbalanced
by the pressure of the poloidal field so that $B'_\phi\approx B'_p$.
Taking into account that $B_{\phi}\propto 1/r$ and $B_{p}\propto 1/r^2$,
this condition immediately implies that the Lorenz factor of the jet is
proportional to the jet's cylindrical radius, $\Gj\approx\rj/\rL$.
In appendix A we show that this scaling could be generalized to the 
non-relativistic case so that generally
\begin{equation}
  \label{eq:ujrjrl}
  \uj\approx \rj/\rL .
\end{equation}
Using this  and $\thj\approx\rj/z$,
the condition  of the strong connection, Eq. (\ref{eq:ujthj}) reduces to $\rj<\sqrt{\rL z}$
which is always fulfilled for our jet. 
The radius of the jet {in the region where it is relativistic}  is
found from the simple relation 
\citep{2008AIPC.1054...71T,2009MNRAS.394.1182K,2009ApJ...698.1570L,
2011PhRvE..83a6302L}
\begin{equation}\label{eq:r_eq}
  \rj = \rL\fracb{\pL}{\pc}^{1/4},
\end{equation}
where  $\pL$ is the jet's magnetic pressure at $\rL$.

\begin{figure}
\includegraphics[width=6in]{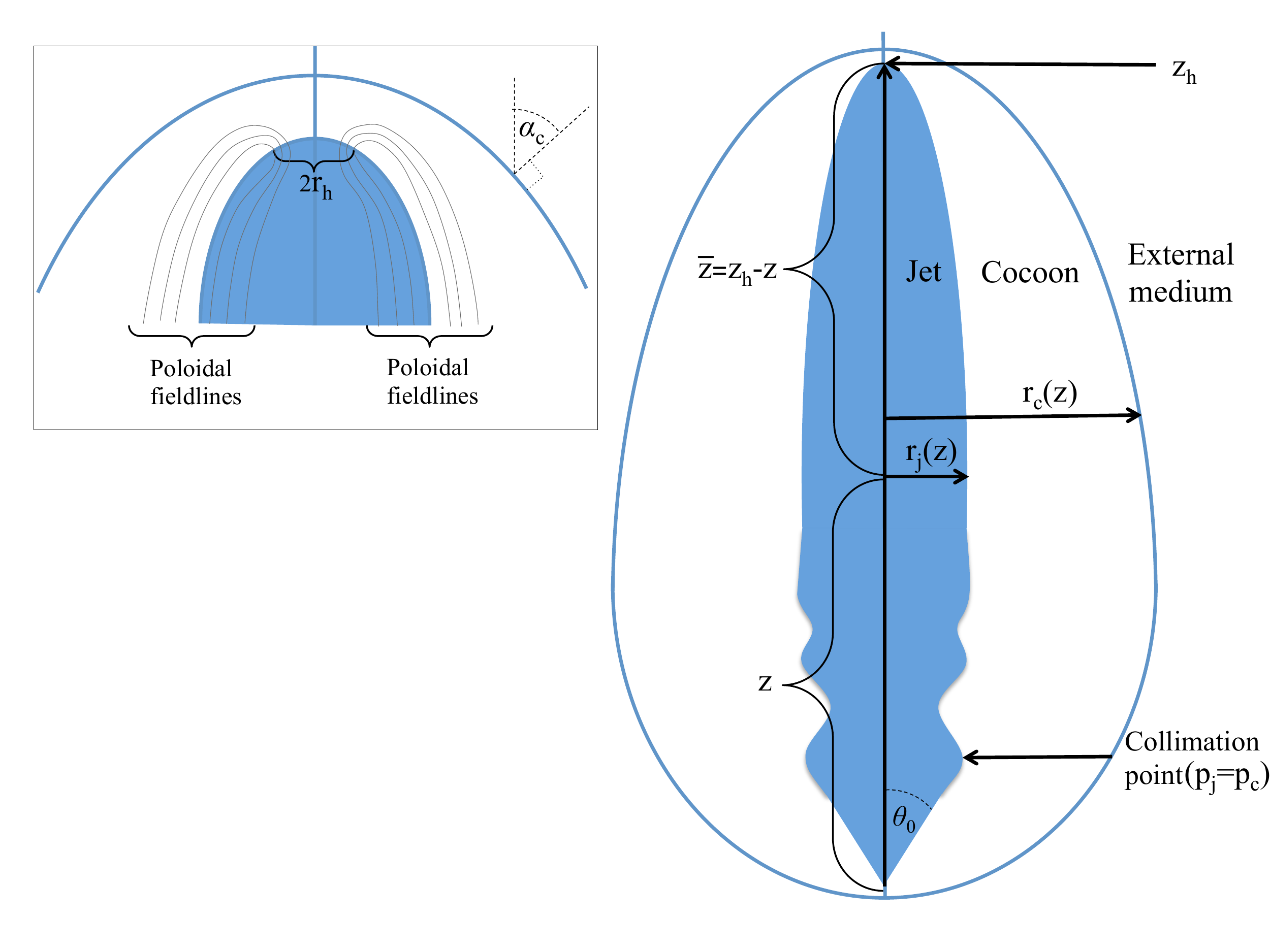}
\caption{The morphology of the magnetic jet and its cocoon.
The entire jet is shown on the right side  and a  zoom in
on the head region is shown on the left.  The jet (blue) is surrounded by a hot cocoon that
applies pressure on it. The cocoon is bounded by a shock surface, that
separates it from the external medium (solid blue line).  Below the
collimation point the jet's pressure ($\pj$) is larger than the
pressure of the cocoon ($\pc$) and the jet  expands along
conical field lines. When $\pj=\pc$, the jet becomes collimated by the
cocoon, and from there on it maintains pressure balance with it. The
collimation is accompanied by oscillations in the jet's radius. As the jet expands
these oscillations 
gradually die out. Close to the head $\pc$ increases and as a result the jet's cross section decreases.  The jet material
decelerates until it's velocity matches  the velocity of the shocked
material ahead of the jet at a radius $\rj\simeq\rhead$.  The dashed horizontal
line separates the relativistic part of the cocoon from the non-relativistic part. The former exists only when the jet's
head moves at relativistic velocities.}
\label{fig:jet}
\end{figure}

\section{The propagation velocity (The velocity of the jet's head)}\label{sec:vh}

When the magnetic fields components are in equilibrium they are comparable
in the comoving frame.  In this case
the proper velocity of the jet's material scales linearly with the jet's cylindrical radius
(see Eq. \ref{eq:ujrjrl}).
Note that $\uj$ is primarily directed along the poloidal magnetic field lines,
which inside the collimated jet can be approximated as the $z$ direction.
As the fast moving material in the jet approaches the jet's head it
must slow down in order to match its velocity to that of the shocked
ambient matter. Since shocks are weak and ineffective in highly
magnetized flows, the matter decelerates gradually as it approaches
the jet's head, where the jet becomes narrower. This is in contrast
with the case of a hydrodynamic jet where there is a strong reverse
shock, across which there is a strong and instantaneous deceleration
\citep[e.g.][]{2011ApJ...740..100B}.

We define the head of the jet as the region where $\alpha_c''$, the angle between the
normal to the bow shock and the $z$-axis in the rest frame of the head, is smaller than 1 
(see Fig.~\ref{fig:jet}). 
In the head, the  
magnetic pressure is balanced by the ram pressure of the ambient
medium.  We can express this pressure balance as:
\begin{equation}\label{eq:p_balance}
\rext c^2\uh^2\cos^2\alpha_c'' \approx \pj\ ,
\end{equation}
where $\uh$ is the proper velocity of the head and $\rext$ is the mass
density of the ambient medium. 
Since the jet is
highly magnetized it's internal pressure can be approximated as
$\pj=B'^2/8\pi\sim B'^2_\phi/4\pi=(B_\phi/\Gamma_{\rm j})^2/4\pi$.  
Taking the jet's dimensionless 3-velocity,
$\bj$,
the electric field is 
$E\simeq \bj B_\phi$,
and the Poynting flux is
\begin{equation}
  {S}  = (c/4\pi)\bj  B_\phi^2.
\end{equation}
The electromagnetic luminosity of the jet is given by
\begin{equation}\label{eq:Lj}
 \Lj = \int 2\pi rdr{S} \approx
(c/4)B_\phi^2\rj^2{\beta}_{\rm j}
\approx \pi \rj^2\Gamma_{\rm j}^2\beta_{{\rm j}}\pj c.
\end{equation}
Substituting  in
Eqs. (\ref{eq:ujrjrl}, \ref{eq:p_balance}) we obtain
\begin{equation}
\rext c^2\uh^2\cos^2\alpha_c'' \sim
\frac{\Lj}{\pi c \rL^2 \Gj\uj^3}\ .
\end{equation}
Near the head,  $\cos\alpha_c'' \sim 1$
(the forward shock is roughly perpendicular to the $z$-axis). Substituting
$\uj\sim\uh$, we get
\begin{equation}\label{a_zh}
\gh\uh^5 = (1+\uh^2)^{1/2}\uh^5\sim \frac{\Lj}{\pi\rext c^3 \rL^2}\ .
\end{equation}

We define the dimensionless quantity
\begin{equation}\label{eq:a}
a\equiv\frac{\Lj}{\pi\rext c^3\rL^2}
=\frac{\pL}{\rext c^2}\approx 1.2\frac{L_{50}}{\rho_4 r_{_{\rm L7}}^2}\ ,
\end{equation}
which represents the ratio between the jet's magnetic pressure at the light
cylinder and the ambient medium's rest mass energy density near the head.
We use here {and elsewhere the notation $q_x \equiv q/10^x$ in c.g.s. units.}
Now, Eqs. (\ref{a_zh}, \ref{eq:a}) yield:
\begin{equation} \label{Eq:u_a}
u_{\rm h} \sim\frac{\rhead}{\rL} \sim
\left\{\matrix{
a^{1/5} \quad (\uh\ll 1) \ ,\cr\cr
a^{1/6} \quad (\uh\gg 1) \ .
}\right.
\end{equation}
Note that for a given jet (i.e. for a particular $\pL\propto\Lj/\rL^2$)
the parameter $a$, which determines the velocity and the cross-section of
the jet's head, depends only on the ambient density near the head.  Thus,
Eq. (\ref{Eq:u_a})
implies that the jet's propagation velocity is determined locally
and it is insensitive to the geometry of the jet below the head.

\begin{figure}
\includegraphics[width=5in]{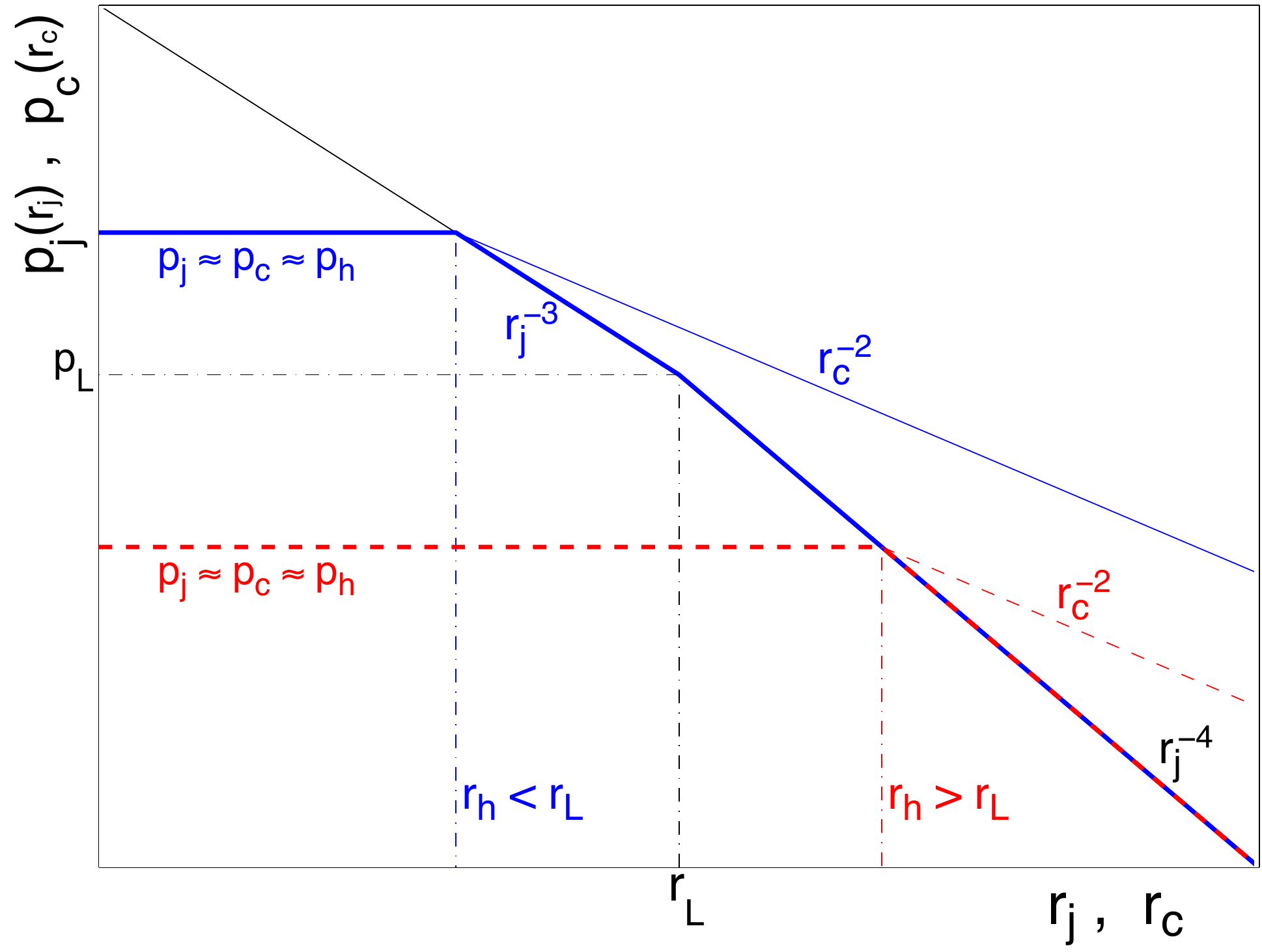}
\caption{The pressure distribution across the jet, $\pj(\rj)$ (Thick line), and the cocoon,
$\pc(\rc)$ (thin line). Solid line shows the distribution in a case of a non-relativistic head, 
and  dashed lines for relativistic head.}
\label{fig:Pjc_rjc}
\end{figure}

\section{The geometry of the cocoon}\label{sec:cocoon}

The energy in the cocoon originates from the work the jet preforms
on the ambient medium.
In the lab frame the differential work is $dE = \pj dV = \pj dA_\perp dz$,
so that the power exerted by the jet is given by $\dot{E}=dE/dt= \int
dA_\perp\vh\pj$, where $\vh = dz/dt$, $A_\perp = \vec{A}\cdot\hat{z}$,
and $dA_\perp = 2\pi\rj d\rj = 2\pi\rj^2 d\log\rj$. Thus the power
per logarithmic jet radius is
\eqb\label{eq:dEdtdlogr}
 \frac{d\dot{E}}{d\log\rj}= \pj \frac{dA_\perp}{d\log\rj}\fracb{dz}{dt}
= \pj 2\pi\rj^2\vh\ .
\eqe
{We  define $\pj $ as a function of $\rj$ in the form (see Appendix A),}
\eqb\label{eq:p_j}
\pj(\rj)  \approx \ph\left\{\matrix{
1 &\quad (\rj < \rhead) \ ,\cr\cr
\fracb{\rj}{\rhead}^{-3} &\quad (\rhead<\rj<\rL)\ ,\cr\cr
\fracb{\rL}{\rhead}^{-3}\fracb{\rj}{\rL}^{-4} &\quad (\rhead<\rL<\rj)\ ,\cr\cr
\fracb{\rj}{\rhead}^{-4} &\quad (\rL<\rhead<\rj)\ ,
}\right.
\eqe
where 
\begin{equation}\label{eq:ph}
  \ph \approx \rext c^2\uh^2
\end{equation}
is the pressure at the jet's
head. The second and third cases in Eq. (\ref{eq:p_j}) are relevant only
when the head is non-relativistic ($\uh\ll 1$), and therefore
$\rhead<\rL$ (see Eq. \ref{eq:ujrjrl}). The fourth case is relevant for a
relativistic head.  Figure \ref{fig:Pjc_rjc} illustrates $\pj(\rj)$ in the different regimes.

A slightly different way of presenting $\pj(\rj) $ 
is by normalizing the pressure to
$\pL=\Lj/(\pi\rL^2c)$, the jet's magnetic pressure at the light
cylinder, which is constant for a given jet.  This gives a sense of
how the head's pressure depends on the ambient medium parameters:
\eqb\label{eq:p_j_P_L}
\pj \approx \pL \left\{\matrix{
\fracb{\rhead}{\rL}^{-4}\simeq a^{-2/3} &\quad (\rj < \rhead>\rL)\ ,\cr\cr
\fracb{\rhead}{\rL}^{-3}\simeq a^{-3/5} &\quad (\rj < \rhead<\rL) \ ,\cr\cr
\fracb{\rj}{\rL}^{-3} &\quad (\rhead<\rj <\rL)\ ,\cr\cr
\fracb{\rj}{\rL}^{-4} &\quad (\max\{\rL,\rhead\} < \rj)\ .
}\right.
\eqe
Here, the first case is relevant for a relativistic head while the
second and third cases are relevant for a non-relativistic head, as before.

Substituting Eq. (\ref{eq:p_j}) or (\ref{eq:p_j_P_L}) into
Eq. (\ref{eq:dEdtdlogr}) we obtain:
\eqb\label{eq:dEdtdlogr_rj}
\frac{d\dot{E}}{d\log\rj} \propto \left\{\matrix{
\rj^2 & \quad (\rj < \rhead) \ ,\cr\cr
\rj^{-1} & \quad (\rhead<\rj<\rL) \ ,\cr\cr
\rj^{-2} & \quad ({\rm max}\{\rL,\rhead\}<\rj)\ .
}\right.
\eqe
Eq. (\ref{eq:dEdtdlogr_rj}) implies that the total $pdV$ work
preformed by the jet on the ambient medium is dominated by the
contribution from $\rj\leq\rhead$.  Therefore, the total rate of energy
injection into the cocoon
can be approximated as
\eqb\label{eq:dotE}
\dot{E}_c \approx \pi \ph\vh\rhead^2\ .
\eqe
It is interesting to view the fraction of the jet luminosity that is
transferred into the cocoon in this way. This  is a measure of the
efficiency of this process. Using Eqs. (\ref{eq:ujrjrl}, \ref{eq:ph})  gives
\begin{equation}\label{eq:EdotL}
\frac{\dot{E}_c}{\Lj} \sim \frac{\pi\rext c^3\rL^2\uh^4\beta_{\rm h}}{\Lj}
\sim \frac{\uh^4\bh}{a} \sim \frac{1}{\gh^2}\ .
\end{equation}
This implies that when the head of the jet is non-relativistic ($\uh\ll1$)
most of the source luminosity is channeled into the
cocoon through the work performed by the head. However when the head
becomes relativistic ($\uh\gg1$) this fraction
decreases by a factor of $1/\gh^2\ll1$. A similar ratio of
$\dot{E}_c/\Lj$ was found in hydrodynamic jets as well
(Bromberg et al. 2011a).

In order to calculate the  cocoon's geometry, as
parameterized by its cylindrical radius $\rc(z)$, we treat separately
the {region where} the cocoon is relativistically hot  ($\pc\gg\rext c^2$),
and {the region where} it is Newtonian ($\pc\ll\rext c^2$).
{The former region exists}
only in jets with $\uh\gg1$. It is characterized by a
relativistic expansion of the cocoon and in the case of a collapsar jet
it is limited to a narrow range near the head of the jet (see
Fig.~\ref{fig:jet}).  The analysis in this region is done under the
approximation that the cocoon's shape is at a steady state in the rest
frame of the head. This assumption is exact in the case when the
ambient medium is uniform. In this case the head propagates at a
constant velocity (see Eq. \ref{a_zh}), and the geometry of the
shock in the relativistic part of the cocoon is 
{constant in time}.  As we show
below, this steady state approximation holds for most situations that
are relevant for collapsar jets.  
{In the Newtonian cocoon region the gas motion is non-relativistic. Here}
we approximate
the expansion  of the cocoon as cylindrical in the frame of
the ambient medium.
{In both cocoon regions we
assume that the pressure is uniform in the ${\bf\hat r}$ direction}.

\subsection{The relativistic cocoon regime}\label{subsec:rel_cocoon}
{We tun now to} examine the rate of production of internal energy at the
bow shock that separates the relativistic cocoon from the ambient medium.
Unlike the usual analysis of relativistic
shocks we preform the analysis in the frame of the star.
We first calculate the flow of rest mass $M$ through the shock per
logarithmic interval of $r_c$. In a time $dt$ the shock covers a
volume $dV_{<\rc}=\pi\rc^2\vh dt$, where $dV_{<\rc}$ is the volume at
radius $<\rc$.  The mass that flows through the shock during the time
interval $dt$ is  $dM_{<\rc}=\rext dV_{<\rc}$ thus:
\begin{equation}\label{eq:dM}
\frac{d^2M}{dtd\log\rc} \equiv \frac{d\dot{M}}{d\log\rc}=
2\pi\rc^2\rext\vh\ .
\end{equation}
The ratio of  energy density to rest mass energy density
downstream of the shock is:
$e/\rho c^2=\bar{\Gamma}$, 
where we define $\bar{u}$ as the proper velocity perpendicular
to the shock in the downstream frame, and $\bar{\Gamma}$ as the Lorentz
factor associate with this velocity \citep[e.g.][]{2006ApJ...651L...1B}. 
The boost to the  lab frame gives
an additional factor of $\bar{\Gamma}$.
Therefore the internal energy per particle of the shocked material, in the lab frame, is
$\tilde{e}_{\rm int}/\tilde{\rho} c^2\equiv e\bar{\Gamma}/\rho c^2-1=
(\bar{\Gamma}^2-1)=\bar{u}^2$. 
The generation
rate of total internal energy, 
$E_{\rm int}$, behind the shock, as measured in the lab frame, is therefore
\begin{equation}\label{eq:Eint}
\frac{d\dot{E}_{\rm int}}{d\log r_c} = 
\fracb{d\dot{M}}{d\log r_c}\frac{\tilde{e}_{\rm int}}{\tilde{\rho}}
\simeq 2\pi r_c^2\rho_{\rm ext}c^2\bar{u}^2v_h \propto r_c^2\bar{u}^2\ .
\end{equation}

In the limit of a strong shock, $\bar{u}$ can also be approximated as the
normal component of the upstream proper velocity in the frame where
the shock is stationary (in our case it is the rest frame of the
head), i.e.:
\eqb\label{eq:ubar}
\bar{u} \simeq u_{\rm h}\cos\alpha_c'' = \frac{u_{\rm h}}{\sqrt{1+\Gamma_{\rm
h}^2\tan^2\alpha_c}}\ .
\eqe
The relativistic cocoon regime corresponds to the regime where $\bar{u}>1$.
It can be seen from Eq. (\ref{eq:ubar}) that in this regime, the cocoon is 
conveniently
separated into two zones
(see fig. \ref{fig:cocoon} for illustration of the different cocoon regions):
\begin{enumerate}
 \item
 The head of the cocoon, defined by the condition $\alpha_c \simless
  1/\gh$ (or $\alpha''_c\simless 1$),   is identified with
  $\rc<\rhead$.
  
  \item 
  The transition region, defined by $1/\gh\simless \alpha_c \simless 1$
  (or $1/\gh \simless \pi/2-\alpha''_c \simless 1$).
  Here $\bar{u} \approx \bh/\tan\alpha_c >1$, hence the bow shock is {still} relativistic.
\end{enumerate}
 
\begin{figure}
  \centering
  
  \includegraphics[width=10cm]{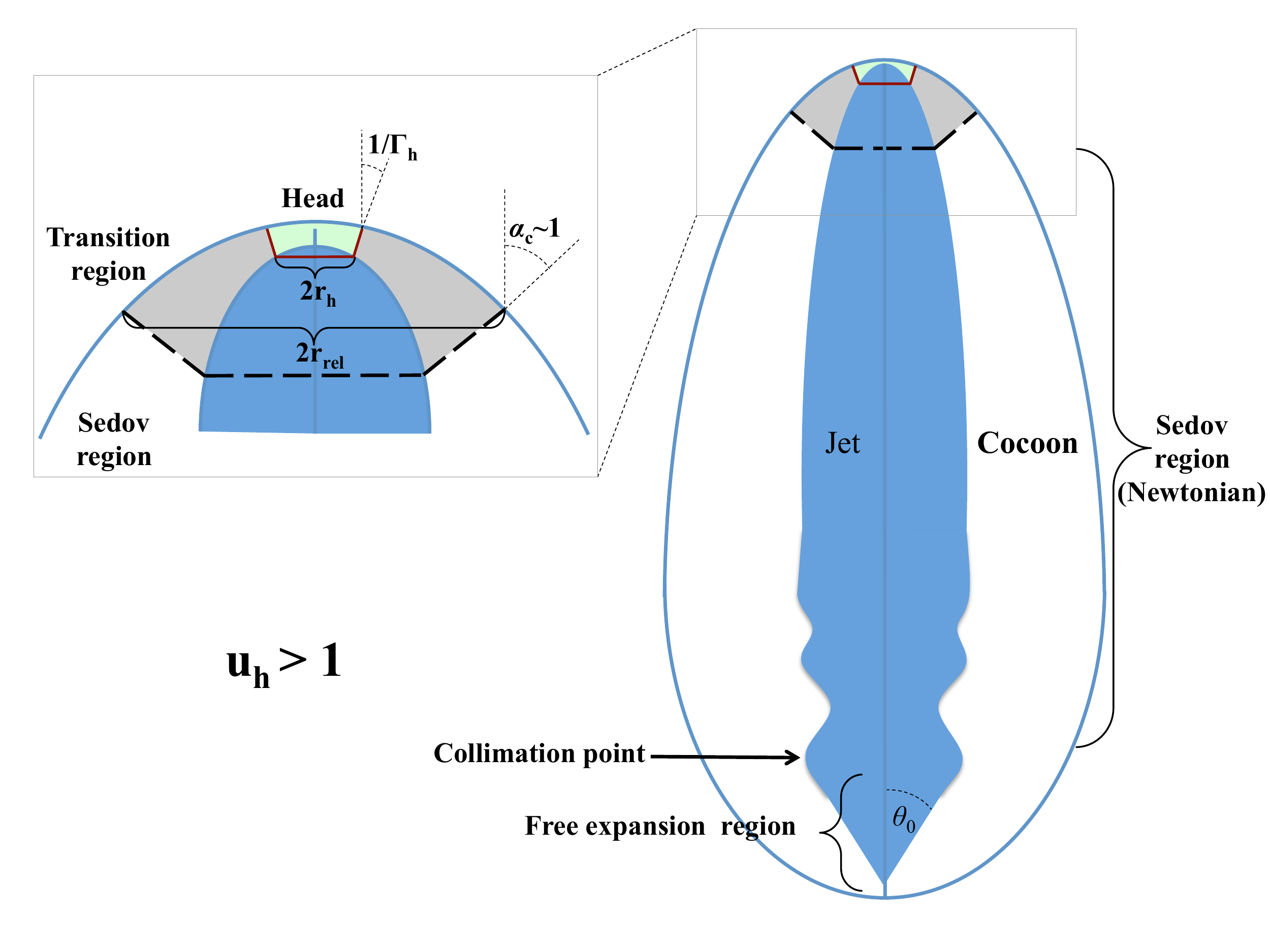}
  \includegraphics[width=10cm]{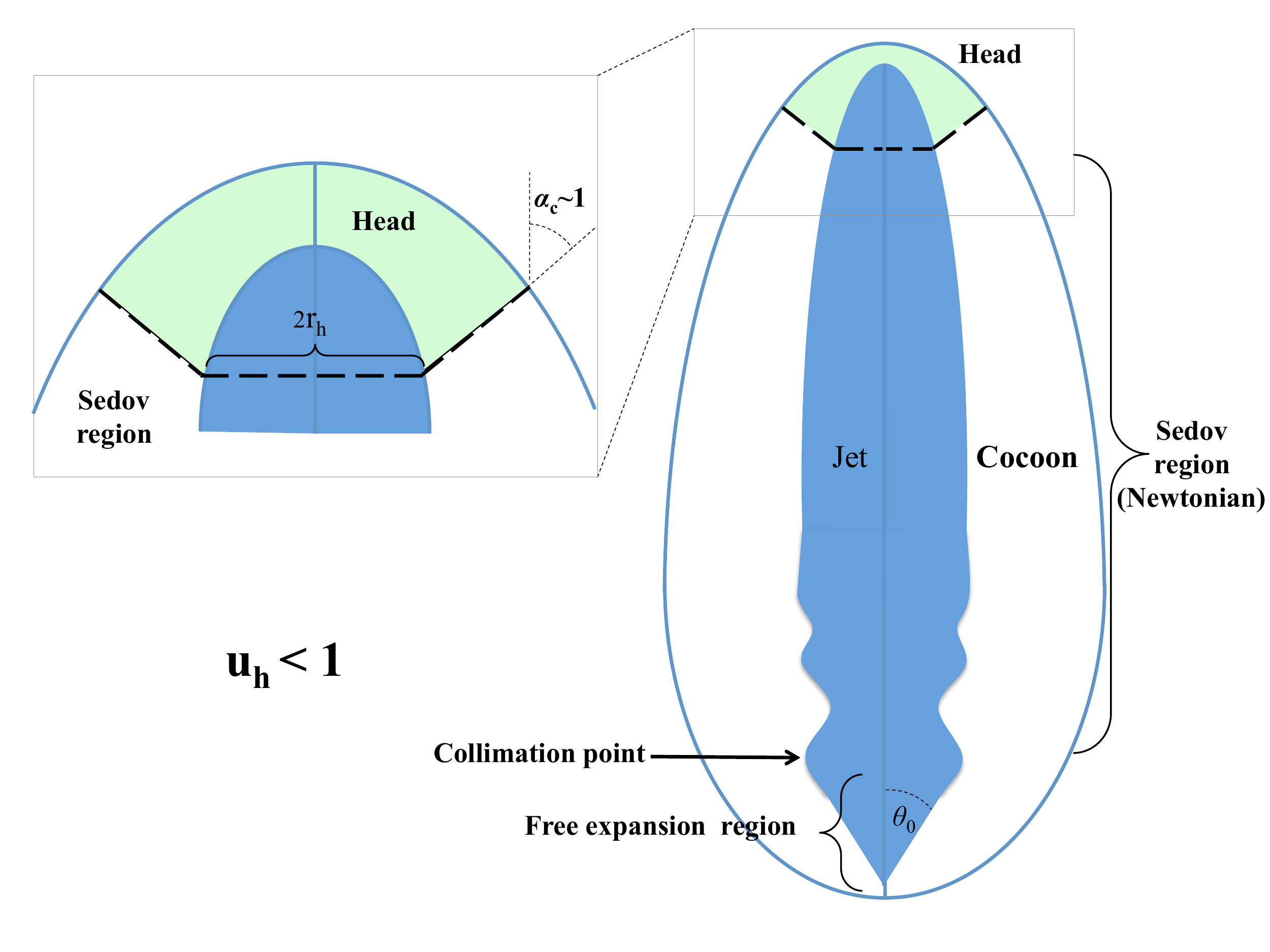} 
  \caption{A schematic description of the different cocoon regions in a case of 
  a relativistic head (top panel) and a non-relativistic head (bottom panel).
  In the relativistic case the cocoon is divided into three regions: a head
  (green), a transition region (gray) where
  the velocities are still relativistic, and a Newtonian region (white),
  which behaves like a cylindrical Sedov-Taylor blast wave.
  In the non-relativistic case the transition region disappears, leaving
  only the head and the Newtonian, Sedov-Taylor, region.
    }
  \label{fig:cocoon}
\end{figure}

  At the cocoon's head the plasma moves upwards at a velocity $\uh$(in the lab frame), 
  and the pressure is roughly uniform, 
  $p_c\approx \rext c^2\bar{u}^2 \approx \rext c^2\uh^2 \approx \ph$.
  We have shown in section 3 that most of the energy is injected into the 
  cocoon in this region.
  This energy is transferred downwards along the cocoon (in the head frame) 
  heating new portions of plasma entering via the wings of the bow shock.
 
   In the transition region the plasma motion is quite involved.
  Close to the head it is still relativistic, however it approaches 
  $\bar{u}=1$ at the interface with the Newtonian cocoon region (see fig. 
  \ref{fig:cocoon}). The geometry of the transition region can be obtained using
  the assumption that it is at a steady state in the head's frame.   
  Consider a sub-region of $r<\rc$ in the transition region. 
  The mass is entering  this sub-region through 
  the bow shock at a rate $\dot{M}_{\rm ext}=\pi \rc^2\rext\bh c$. 
  The energy is entering this sub-region through the interface
  with the head (see fig. \ref{fig:cocoon}) at a rate $\dot{E_c}\simeq L/\gh^2$
  (Eqs. \ref{eq:dotE}, \ref{eq:EdotL}).
  Since the flow is at a steady state,
  mass and energy are flowing out, at $\rc$, in the same rate.
  Now, since in the lab frame  
  $\dot{E_c}\sim \dot{M_{\rm ext}}c^2\bar{u}^2$ (e.g. Eq. \ref{eq:Eint}),
  it implies that
  $\bar{u}^2\sim \dot{E_c}/\dot{M_{\rm ext}}c^2\sim L_j/(\pi\rext c^3 r_c^2 \Gamma_h\uh)$.
 Using Eqs. (\ref{eq:a}, \ref{Eq:u_a}) we
  get that this velocity is equal to
\begin{equation}\label{eq:bar_u} 
\bar u\approx u_{\rm h}\rhead/r_c.
 \end{equation}
  On the other hand, $\tan\alpha_c=d\bar z/dr_c$, where
  $\bar z\equiv z_h-z$ is the distance from the jet's head in the lab
  frame (see Fig.~\ref{fig:jet}), implying that $\bar u \approx
  \beta_h/\tan\alpha = \beta_h{dr_c}/{d\bar z}$. Assuming a functional
  shape of $r_c = \bar z^\epsilon$ we obtain that
\begin{equation}\label{eq:ubar1}
\bar{u}\approx\epsilon\bh\rc/\bar{z}\ .
\end{equation}
  Equating Eq. (\ref{eq:ubar1}) with Eq. (\ref{eq:bar_u}) provides the
  profile of $\rc$ in the transition region,
\begin{equation}\label{eq:r_c}
\rc \approx\sqrt{2\rhead\gh\bar{z}}\ ,
\end{equation}
  and gives $\epsilon=1/2$.

The pressure in the cocoon , $\pc(\bar{z})$, which we assume to be
nearly independent of $r$, can be estimated just behind the shock: 
\begin{equation}\label{eq:p_c}
\pc \approx \rext c^2\bar{u}^2 = \rext c^2 \left(\frac{\bh\rc}{2\bar{z}}\right)^2\ ,
\end{equation}
  where we used Eq. (\ref{eq:ubar1}) for $\bar{u}$.  Substituting $\rc$
  from Eq. (\ref{eq:r_c}) we obtain:
\begin{equation}\label{eq:p_c2}
\pc \approx \ph \frac{\rhead}{2\gh\bar{z}} = \ph\bh\frac{\rL}{2\bar{z}}\ ,
\end{equation}
  where we use the relation $\ph=\rext c^2\uh^2$.  The jet radius
  is calculated using $\rj=\rhead(\ph/\pc)^{1/4}$ (Eq. \ref{eq:r_eq}), resulting in:
\begin{equation}\label{eq:r_j}
\rj\approx \rhead\left(\frac{2\bar{z}}{\bh\rL}\right)^{1/4}\ ,
\end{equation}
 Note that $\bar{u}\propto r_c^{-1}$ as we show in Eq. (\ref{eq:ubar})
 implies that $d\dot E_{\rm int}/d\log r_c={\rm const}$,
 (Eq. \ref{eq:Eint}). This means that  
 most of the energy, which was injected by the jet at $r_c\simless\rhead$ is given to newly 
 shocked material each dynamical time. Such a behavior  is expected in an adiabatic 
 (i.e. constant energy) blast wave solution, like the Blandford-McKee or Sedov-Taylor  
 solutions. 
 In both cases $E\sim M\bar{u}^2$ and $M\propto \rc^2$.
This type of
  solutions apply also to the regime where the shock becomes
  Newtonian, as we show in the next section.

{We can now} examine the assumption of a uniform $\rext$
  in the relativistic cocoon regime. The transition point from a relativistic
  shock to a Newtonian shock occurs when $\bar u = 1$. From
  Eqs. (\ref{eq:bar_u}, \ref{eq:ubar1}) we obtain that this occurs when
\begin{equation}\label{eq:barzrel}
\bar{z}\equiv\bar{z}_{\rm rel} \approx \frac{1}{2}\rL\uh^2\bh\ 
\quad{\rm and} \quad
\rc \equiv r_{\rm rel} \approx \rL\uh^2 \approx 2z_{\rm rel}\ .
\end{equation}
  For any reasonable values of jet luminosity and stellar properties
  the Lorentz factor of the jet's head remains $\uh\lesssim 10$, until
  it breaks out of the stellar surface. Taking a reasonable estimation
  for the light cylinder radius, $\rL\simeq 10^7\;$cm, results in a
  relativistic region of size $\bar{z}_{\rm rel}\lesssim 10^9\;$cm
  which is about two orders of magnitude smaller than the size of the
  jet before it breaks out of the star.  In addition, the density of
  the stellar envelope changes on length scales much larger than
  $10^9\;$cm at this region (e.g. Mizulta \& Alloy 2009).  Deeper in
  the star the size of the relativistic region decreases roughly
  linearly with $z_h$ for typical density profiles, and it remains
  smaller than $\rext/|\nabla\rext|$ throughout the entire star. Thus
  the approximation of a constant ambient medium density at the
  relativistic cocoon region is justified.

\subsection{The Newtonian cocoon regime}\label{subsec:non-rel_cocoon}

When $\alpha_c\simgreat 1$ ($\alpha''_c\simgreat\pi/2-1/\gh$)\footnote{{The
condition $\alpha_c\simgreat 1$ is relevant also in the case when the head velocity
is sub relativistic. In this case the region with $\alpha_c\simless 1$ is occupied by the head
(see Fig. \ref{fig:cocoon}).}} , 
the shock becomes Newtonian ($\bar{u}<1$). 
This region covers most of the jet
while it is still inside the star. 
{When the head of the jet is sub-relativistic it engulfs the entire jet 
(see Fig. \ref{fig:cocoon}).}
Here we cannot ignore the density
gradient in the stellar envelope and thus the approximation of a
steady state shock doesn't apply. Namely, while the relation
$\pc\approx \rext c^2\bar{u}^2$ still holds, Eq. (\ref{eq:ubar}) can
no longer be used to evaluate $\bar{u}$ (the relative proper velocity
between the upstream and the downstream).  
Instead we use the assumption that all motions in this region are in the $\bf\hat{r}$
direction, and that the expansion follows a cylindrical Sedov-Taylor solution,
a low energy extension to the cylindrical Blandford-MaKee solution that was
used in the relativistic cocoon region.
In this type of expansion the cylindrical radius follows:
\begin{equation}\label{eq:Delta_rc}
  r_c^2(z)={{2}}\sqrt{\frac{\mathcal{E}(z)}{\pi\rext(z) }}\Delta t,
\end{equation}
where 
 $\mathcal{E}(z)$ is the energy per unit length that was injected by the head when it
 passed through altitude $z$ (i.e. when $\zh=z$):
\begin{equation}\label{eq:Eepsilon_z}
\mathcal{E}(z)\equiv\frac{dE_c(z)}{dz}=\frac{\dot{E}_c(z)}{\vh(z)}
\simeq\frac{L_{j}}{\Gamma_h(z)u_h(z)c}\ ,
\end{equation}
and we used   Eqs. (\ref{eq:dotE}, \ref{eq:EdotL}) to express it in terms of  the jet luminosity. 
Hereforth we use the terms $\gh(z)$
,$\uh(z)$ and $\vh(z)=c\bh(z)$ to express the head's Lorentz factor, proper velocity
(i.e. the spatial part of the 4-velocity) and 3-velocity respectively, when the 
head passes altitude $z$
(this terminology also applies below to $a(z)$,
$z_{\rm rel}(z)$ and $r_{\rm rel}(z)$).
The numerical coefficient $2/\sqrt{\pi}\simeq 1$, in Eq. (\ref{eq:Delta_rc}) was chosen to 
obtain a smooth transition
between the relativistic and the Newtonian cocoon regions.

To calculate
$\rc(z)$ we need to evaluate $\Delta t$, the time that is available for the 
cocoon to expand since the energy is injected when the head crosses altitude 
$z$, until it reaches $\zh(t)$.  We can use the substitution 
 $dt\equiv d\zh/\bh(z)c$ and integrate over the propagation time of the head
 between the two limits. 
 In order to obtain useful analytic
expressions we use  a power-law external density profile,
$\rext\propto z^{-\xi}$ with $\xi\geq 0$, and denote by $z_1$  the
altitude where the jet's head becomes relativistic, namely
$\uh(z_1)\equiv 1$, so that to a zero order 
\eqb\label{eq:beta_h_z}
\bh(z) = \left\{\matrix{
a(z)^{1/5}= (z/z_1)^{\xi/5}
& \quad (z<z_1)\ ,\cr\cr
1 & \quad (z>z_1)\ .
}\right.
\eqe
Note that this also implies $\uh(z)\simeq\min[(z/z1)^{\xi/5},
(z/z1)^{\xi/6}]$.

We distinguish between
three cases: a non-relativistic head ($z<\zh<z_1$), a relativistic  head at $z$ ($z_1<z<\zh$),
and a case where the head is non relativistic at altitude $z$ but becomes relativistic
before it reaches $\zh$ ($z<z_1<\zh$). This gives:
\eqb\label{eq:dt}
c\Delta t = \int_{z}^{z_h(t)}\frac{dz}{\beta_h(z)} = \left\{\matrix{
\frac{5}{5-\xi}\left(\frac{z_h(t)}{\beta_h(z_h)}-\frac{z}{\beta_h(z)}\right) &
\quad (z<\zh<z_1)\ ,\cr\cr
\frac{5}{5-\xi}\left({z_1}-\frac{z}{\beta_h(z)}\right)+z_h(t)-z_1 &
\quad (z<z_1<\zh)\ ,\cr\cr
z_h(t)-z & \quad (z_1<z<\zh)\ ,
}\right.
\eqe
Altogether the solution to Eq. (\ref{eq:Delta_rc}) is
\eqb\label{eq:r_c_NR}
r_c^2(z,t) = 2\rL
\left\{\matrix{
\fracb{z}{z_1}^{2\xi/5}
\frac{5}{5-\xi}\left(\frac{z_h(t)}{\beta_h(z_h)}-\frac{z}{\beta_h(z)}\right) &
\quad (z<\zh<z_1)\ ,\cr\cr
\fracb{z}{z_1}^{2\xi/5}\left[\frac{5}{5-\xi}\left({z_1}-\frac{z}{\beta_h(z)}\right)
+z_h(t)-z_1\right] & \quad (z<z_1<\zh)\ ,\cr\cr
\fracb{z}{z_1}^{\xi/3}\left[z_h(t)-z\right] & \quad (z_1<z<\zh)\ .
}\right.
\eqe

The cocoon pressure is evaluated as $p_c(z,t)\simeq \mathcal{E}(z)/\pi r_c^2(z,t)$, 
where we use the assumption that the pressure is roughly uniform in the $\bf\hat{r}$
direction
\begin{equation}\label{eq:p_c_NR}
p_c(z,t)\simeq\frac{\rext(z)c^2\uh(z)^2}{2}\rL\left\{\matrix{
\frac{5-\xi}{5}\left(\frac{\zh(t)}{\bh(\zh)}-\frac{z}{\bh(z)}\right)^{-1} &
\quad (z<\zh<z_1) \ ,\cr\cr
\left[\frac{5}{5-\xi}\left({z_1}-\frac{z}{\beta_h(z)}\right)+z_h(t)-z_1\right]^{-1} &
\quad (z<z_1<\zh) \ ,\cr\cr
\left(\zh(t)-z\right)^{-1} & \quad (z_1<z<\zh)\ .
}\right.
\end{equation}
Note that $\rext(z)c^2\uh(z)^2$ is just the pressure at the
jet's head when it passes the altitude $z$.  The jet's radius
satisfies $\rj=\rL(\pL/\pc)^{1/3}$ when $\rj<\rL$ and
$\rj=\rL(\pL/\pc)^{1/4}$ when $\rj>\rL$.  Substituting $\pc$
from Eq. (\ref{eq:p_c_NR}) and using $\pL=\Lj/(\pi\rL^2c)$ we obtain
\begin{equation}\label{eq:r_j_NR}
\rj(z,t)=\rL \left\{\matrix{
\fracb{z}{z_1}^{\xi/5}\fracb{2\kappa}{\rL}^{1/3}
\left(\frac{\zh(t)}{\bh(\zh)}-\frac{z}{\bh(z)}\right)^{1/3} &
\quad (z<\zh<z_1; \rj<\rL) \ ,\cr\cr
\fracb{z}{z_1}^{3\xi/20}\fracb{2\kappa}{\rL}^{1/4}
\left(\frac{\zh(t)}{\bh(\zh)}-\frac{z}{\bh(z)}\right)^{1/4} &
\quad (z<\zh<z_1; \rj>\rL) \ ,\cr\cr
\fracb{z}{z_1}^{3\xi/20}\fracb{2\kappa}{\rL}^{1/4}
\left(z_1-\frac{z}{\beta(z)}+\frac{\zh(t)-z_1}{\kappa}\right)^{1/4} &
\quad (z<z_1<\zh)\ ,\cr\cr
\fracb{z}{z_1}^{\xi/6}\fracb{2\kappa}{\rL}^{1/4}\left(\zh(t)-z\right)^{1/4} &
\quad (z_1<z<\zh)\ .
}\right.
\end{equation}
where $\kappa=5/(5-\xi)$ when $z<z_1$ and $\kappa=1$ when
$z>z_1$.
 Eqs. (\ref{eq:r_c_NR} -- \ref{eq:r_j_NR}) describe the {geometries }
 of the cocoon and of the jet in the  Newtonian region.
 
 {To illustrate the jet's and the cocoon's shape in our model we plot  
 in fig. \ref{fig:jet_analytical} the results of $\rj(z,\zh)$ and $\rc(z,\zh)$, 
 calculated with four different density profiles. To demonstrate the temporal
 evolution of the jet in each profile, we preform the calculation at
 $\zh/z_1= 0.3, 1, 3$. In all cases we use $\rL=0.01 z_1$.
 In fig. \ref{fig:jet_simulation} we show a comparison of our
 model to a numerical simulation of a Poynting dominated collapsar jet
 \citep{BT14}.
 The figure shows snapshots of density (left panel), proper velocity (middles panel) and the
pressure (right panel) taken 2 sec after the initiation of the jet. In each panel
we plot in gray lines our analytic calculation of $\rj(z,t)$ (Eqs. \ref{eq:r_j}, \ref{eq:r_j_NR}) and 
$\rc(z,t)$ (Eqs. \ref{eq:r_c}, \ref{eq:r_c_NR})
at the same time of the snapshots. The radius of the simulated jet is 
best observed in the proper velocity panel, since $\uj$ is expected to reach
its peak value close to the interface with the cocoon.
The analytically calculated $\rj$ tracks the location of this peak 
remarkably well. There is also a very good match in $\pc$and $\rc$
between the simulation and our model.
 The simulation and the analytic calculation were conducted using
 a star with $M=15 M_\odot$, $R=10^{11}$ cm
 and a powerlaw density profile, $\rext\propto z^{-2.5}$. The jet parameters
 are $L_j=10^{50}$ erg/s and $\rL=5\times10^7$ cm.    
 The full details
 of the comparison are given in \citet{BT14}.}

\begin{figure}
  \centering
  \includegraphics[width=15cm]{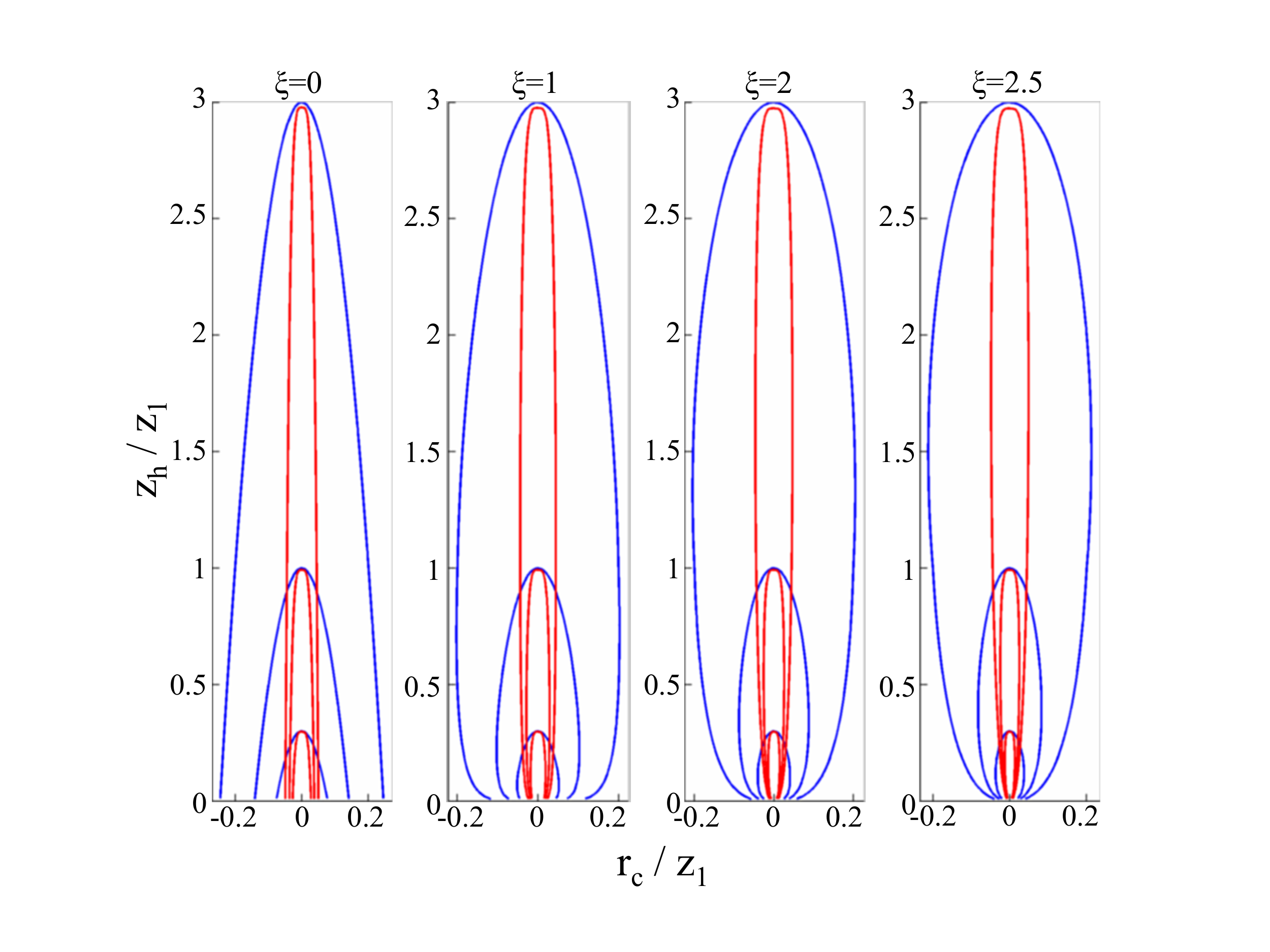}
  \caption{{A schematic drawing of the jet (red) and the cocoon (blue) in four 
  different density profiles: $\rext\propto z^{-\xi}$, and in three altitudes:
  $\zh/z_1=0.3,1,3$. 
  Length scales are normalized to $z_1$, and we chose $\rL/z_1=0.01$.}
  }
  \label{fig:jet_analytical}
\end{figure}

\begin{figure}

  \centering


  \begin{tabular}{ccc}


    \includegraphics[width=50mm, height=100mm]
    {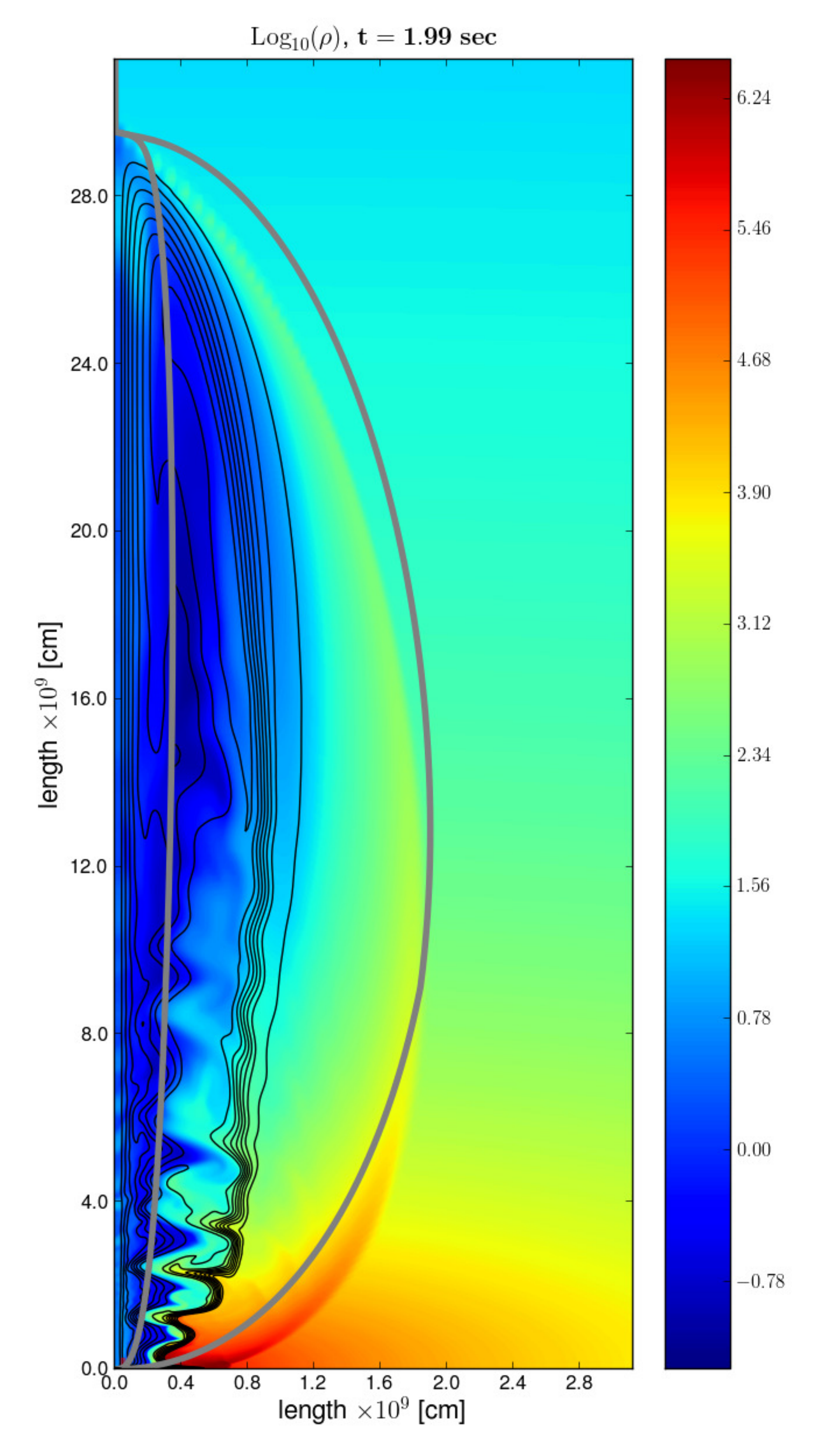}&

    \includegraphics[width=50mm, height=100mm]
    {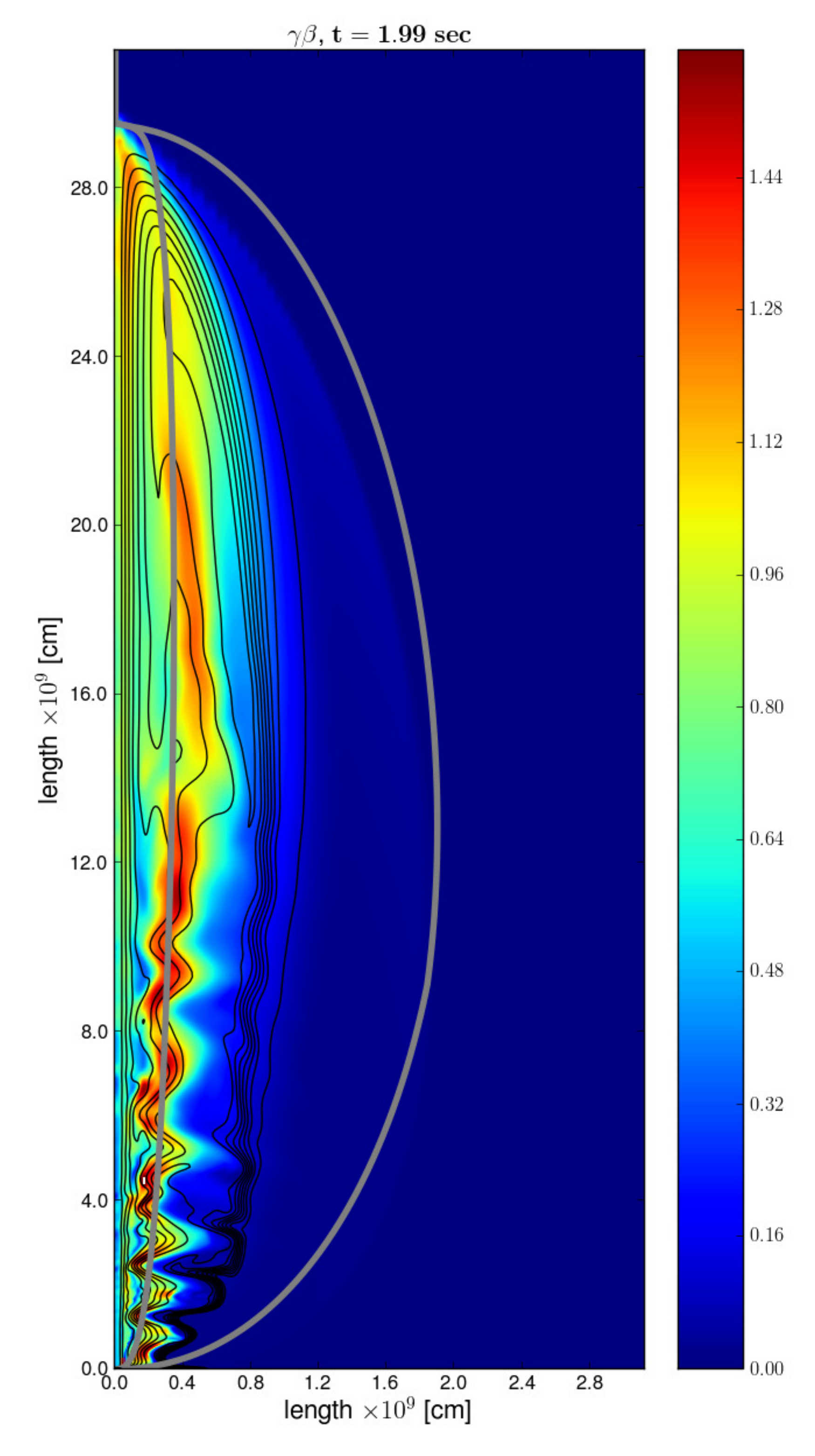}&

    \includegraphics[width=50mm, height=100mm]
    {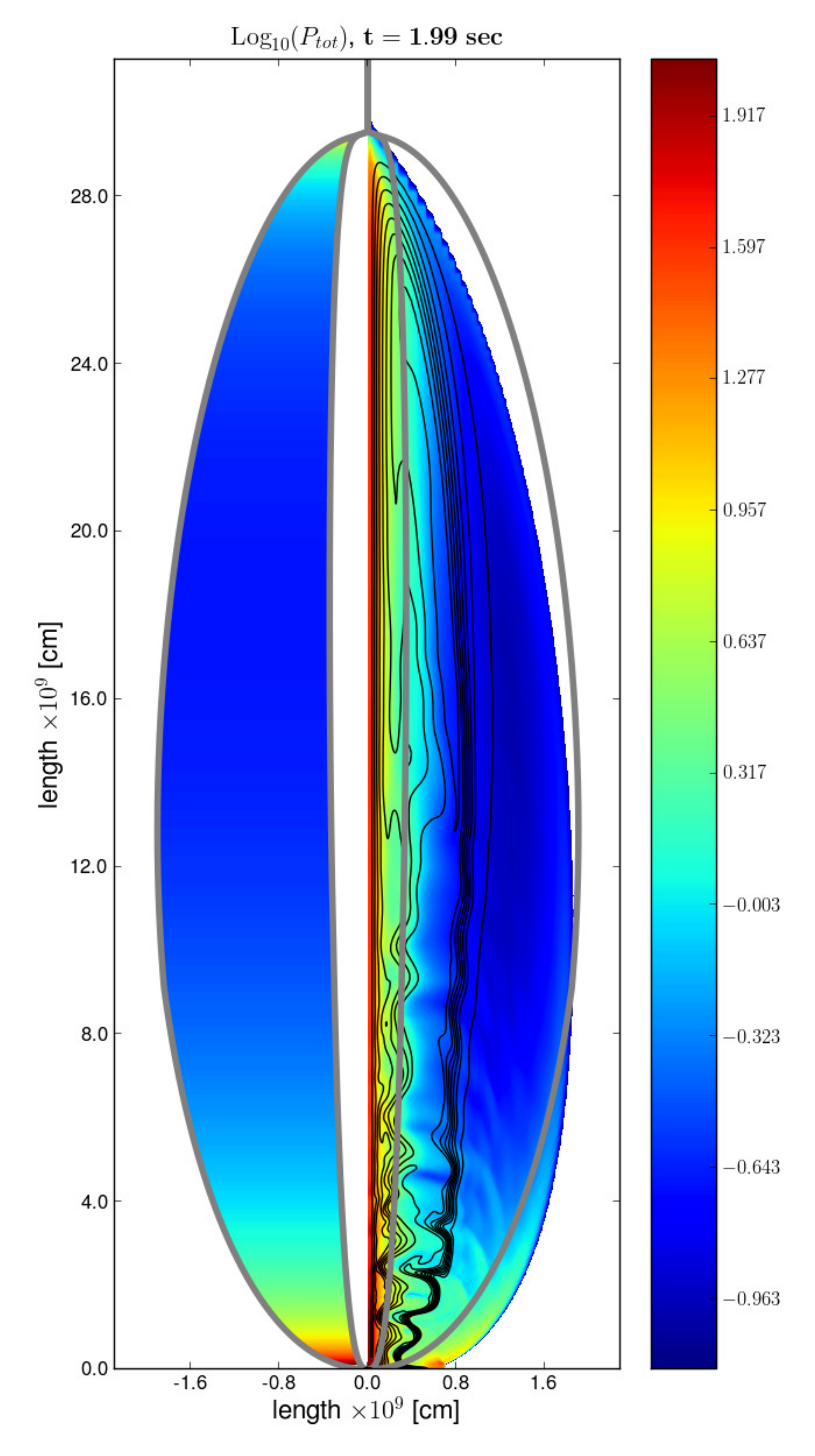}\\

   \end{tabular}
\caption{{A comparison of the analytic model to a 2D simulation
    of a Poynting dominated collapsar jet from \citet{BT14}. 
    We show snapshots taken 2 sec
    after the jet is initiated. Left panel depicts $\log_{10}\rho$ in units of
    ${\rm g/cm^3}$; middle panel: $u =\Gamma\beta$; and right
    panel: $\log_{10}(p)$ in units of $10^{20}~{\rm erg/cm^3}$.
    In the right, pressure, panel we show the simulated pressure on the right  and
    the analytic calculation of $\pc(z,t)$ on the left.
    The thick gray lines mark the analytic values of $\rj(z,t)$ (inner line) and $\rc(z,t)$ (outer
    line) calculated at the same time as the snapshots.    
    The heights of the analytic results are
    normalized by a factor of 1.3 to match the hight of the
    simulated jet. For further details see \citet{BT14}} } 
    \label{fig:jet_simulation}
\end{figure}

\section{The collimation of the jet at the base}\label{sec:jet base}

When the jet pushes its way through the stellar material, the cocoon pressure is
generally large enough to confine the flow into a narrow channel such
that the condition of strong connection is fulfilled.
This  can be checked by verifying that the condition (\ref{eq:ujthj})
is fulfilled.
Close to the injection point the jet's pressure is larger than
the cocoon's pressure.  Thus, the jet initially expands freely until it reaches
a point where its
internal pressure matches the pressure in the cocoon.
From this point on the jet  becomes collimated (see Fig.~\ref{fig:jet}).
The behavior of the jet above the collimation point depends on
{the causal connection across the jet } 
at the collimation point (see section \ref{sec:general}).
If causal contact {between the outer part of the jet and the axis} is still maintained at that point, the collimation
is expected to occur without shocks or dissipation of energy.
In this situation the jet energy remains dominated by Poynting flux
all the way to the head.

{ We turn now to } review the conditions near the foot of a collapsar's jet
and determine the circumstances under which
the cocoon's pressure is sufficient to collimate the jet while
the jet still maintains causal connection with the axis.
Suppose that {the central engine produces} 
 a Poynting dominated jet { with a (moderate)} Lorentz factor 
 $\Gamma_{0}{\approx1}$, a
magnetization  $\sigma_0\gg1$ and
with an opening angle {$\theta_j\ge\Gamma_0^{-1}$}.  
{Close to the central engine}, the pressure of the jet exceeds
the pressure in the cocoon, thus the flow expands freely.  
{Such an unconfined flow expands
radially, preserving the initial angle. It accelerates linearly up to
the Lorentz factor $\Gamma_{cr}\sim
(\sigma_0/\theta_j^2)^{1/3}$, after which it looses the causal
contact and practically stops accelerating 
\citep{1998MNRAS.299..341B,2010NewA...15..749T}. This occurs at the distance }
\begin{equation}
  z_{\rm cr}\equiv \rL\Gamma_{\rm
  cr}/\thj=\rL\sigma_0^{1/3}\thj^{-5/3}\ .
\end{equation}
Above this point the jet material is no longer causally connected with the axis.  
Therefore the condition for  the collimation to occur at $z\leq z_{\rm cr}$ is
\begin{equation}\label{eq:p_z_cr}
  \pj(z_{\rm cr})\leq \pc(z_{\rm cr}).
\end{equation}

The magnetic pressure of the jet  at $z_{\rm cr}$ is obtained by taking the
condition (\ref{eq:r_eq}) and substituting $\rj=z_{\rm ct}\theta_{\rm j}$:
\begin{equation}\label{eq:p_j_crit}
  \pj (z_{\rm cr}) \simeq \pL\left(\frac{\rL}{\theta_{\rm j}z_{\rm cr}}\right)^{4}=
  \frac{\Lj}{\pi\rL^2c}\sigma_0^{-4/3}\theta_{\rm j}^{8/3}\ ,
\end{equation}
where $\pL = L_{j}/(\pi \rL^2c)$ is the jet's pressure at the light
cylinder.  
A lower limit on the cocoon's pressure at $z_{\rm cr}$ can be calculated by taking  
Eq. (\ref{eq:p_c_NR}) in the limit where $\uh(z)>1$ (assuming $z_{\rm cr}<z_1<\zh$):
\begin{equation}\label{p_c_crit}
  p_c(z_{\rm cr}) \simeq\frac{\rext(z_{\rm cr})c^2\uh(z_{\rm cr})^2\rL}{2}
  \left[\zh-\frac{5}{5-\xi}\frac{z_{\rm cr}}{\bh(z_{\rm cr})}-\frac{\xi}{5-\xi}z_1\right]^{-1}
\end{equation}
Comparison of $\pj(z_{\rm cr})$ with $\pc(z_{\rm cr})$ and requiring the 
condition (\ref{eq:p_z_cr}) yields  a lower limit on the
value of the magnetization, $\sigma_0$:
\begin{equation}
  \sigma_0^{4/3}\geq \frac{2a(z_{\rm cr})}{\uh^2(z_{\rm cr})\rL}
  \left[\zh-\frac{5}{5-\xi}\frac{z_{\rm cr}}{\bh(z_{\rm cr})}-\frac{\xi}{5-\xi}z_1\right]
  \theta_{\rm j}^{8/3}
\end{equation}
{Substitution of $z_h=R$, the stellar radius,  ($R\gg [z_{\rm cr}/\bh(z_{\rm
cr}),z_1]$) in the above expression  gives the minimal value of $\sigma_0$ 
for which a jet, prior to its
breakout, is collimated while it is still causally connected. } 
 Approximating the stellar density as $\rho\propto z^{-\xi}$
we obtain that the minimal value of $\sigma_0$ is
\begin{equation}
  \sigma_0^{4/3-\xi/5} \geq \aL^{3/5}\frac{R}{\rL}\theta_0^{8/3-\xi}
 = 5 
  \left[\frac{0.5}{3-\xi}L_{50}\fracb{15M_\odot}{M}R_{11}^{(14/3-\xi)}r_{L,7}^{(\xi-11/3)}\right]^{3/5}
  \theta_{\rm j}^{8/3-\xi},
\end{equation}
where $\aL\equiv a(z=\rL)$, $M$ is the mass of the star and $R$ is its radius.
For a typical value of $\xi=2.5$ we get that $\sigma_0\simgreat10$.
The  values of $\sigma_0$ that are inferred in GRB jets, if they are
dominated by Poynting flux, are typically above a few times $10^2$, at least
an order of magnitude larger than this limit. With these values
 the jet is collimated smoothly prior to its breakout, without
formation of internal shocks.

\section{Stability of the jet}\label{sec:stability}



Our axisymmetric model implicitly assumes that the jet is not
destroyed by global MHD instabilities.
For a narrow jet, the most dangerous instability is the 
kink instability.  This instability excites
large-scale helical motions that can strongly distort or even disrupt
the jet, thus triggering violent magnetic dissipation 
\citep{1992SvAL...18..356L,1993ApJ...419..111E,1997MNRAS.288..333S,
1998ApJ...493..291B,1999MNRAS.308.1006L,2006A&A...450..887G}.
The instability can develop only in strongly causally
connected flows, like a collimated jet. Therefore the { Poynting
dominated} jet can survive the crossing of the star only if the
crossing time is shorter than the growth time of the instability in
the lab frame. 

The kink instability is sensitive to the structure of the magnetic field
in the flow.  Flows with a purely toroidal field are mostly sensitive
to disruption by this instability 
\citep{2010MNRAS.402....7M,2012MNRAS.422.1436O}.
Such a configuration could however be maintained in a
strongly causally connected flow, only if the hoop stress is balanced by
the plasma pressure and/or by the centrifugal force. This is possible
only when $\sigma\leq 1$. 
Such a situation is relevant for pulsar wind nebulae, where  the kink instability 
plays indeed a crucial role 
\citep{1998ApJ...493..291B,2014MNRAS.438..278P}.
In a strongly connected Poynting dominated
flow, as in our case, in the comoving frame the toroidal field is
comparable to the poloidal field \citep{2009ApJ...698.1570L}.
 {This results in a more stable flow.}

In the simplest case of a,
non-rotating, rigidly moving jet, the electric field vanishes in the
comoving frame, and the hoop stress is balanced by the magnetic
pressure. Thus the poloidal field decreases outwards. In this case, a
characteristic time for the growth of a perturbation is about a few
Alfv´en radial crossing times in the comoving frame. For example, 
\citet{2000A&A...355..818A}
found the growth rate $\kappa=0.133 v_A/r_j$ for a jet with
$B_z(r)=B_0r_j^2/(r_j^2+r^2)$, where $v_A$ is the Alfv'en velocity,
which in our case could be taken to be equal to the speed of
light. The full development of the instability takes generally a few
characteristic growth times. One can estimate the
total time necessary for the full development of the instability as
$t'_{\rm kink} \sim 10f/\kappa\approx{100fr_j/c}$, where $f$
is a numerical factor having a value $f\sim 0.5-1$.  This estimate agrees
with the results of numerical simulations 
\citep{2009ApJ...700..684M,2012ApJ...757...16M}.
The unstable perturbation propagates with the plasma of the jet therefore 
in the observer frame, the instability disruption time is
estimated as
\begin{equation}\label{eq:t_kink}
t_{\rm kink} = \Gj t'_{\rm kink}\sim 100 f\frac{\rj^2}{c\rL} \ ,
\end{equation}
where we used Eq. (\ref{eq:ujrjrl}) to convert $\Gj$. 
{This time should be compared with the dynamical time  that is available
for the instability to grow in the jet.}
{The relevant dynamical time, $t_{\rm dyn}$, is the smaller between the propagation times of a 
fluid element to the jet's head and to the star's edge ($\approx R/c$).}


When the head is Newtonian ($\uh\ll1$) a  fluid element crosses the jet at a time that is
much shorter than the time it takes the jet to double its height.  Therefore
we can treat the jet as if it is stationary during the time it takes
the fluid element to reach the head. Namely the fluid element tracks
the time independent geometry of the jet, $\rj(z,\zh)$. To estimate the height of 
the widest point of the jet
we take the derivative of $\rj(z,\zh)$ in Eq. (\ref{eq:r_j_NR}) with respect to z, 
keeping $\zh$ constant:
\begin{equation}
  \left(\frac{\partial r_j}{\partial z}\right)_{\zh}= \frac{r_j}{z} \left[\frac{3\xi}{20} - 
  \frac{5-\xi}{20}\left(\frac{z^{(5-\xi)/5}}{\zh^{(5-\xi)/5}-z^{(5-\xi)/5}}\right)\right].
\end{equation}
The jet's radius,$\rj(z,\zh)$, has an extremum at:
\begin{equation}\label{eq:bar_z_max_NR}
  {z}_{\rm max}\equiv z({\partial\rj}/{\partial z}=0)=0.56 \zh
 \fracb{3\xi}{0.75(5+2\xi)}^{\frac{5}{5-\xi}}.
\end{equation}
A fluid element that is injected at the base of the jet expands as long as $z<z_{\rm 
max}$. To estimate the growth of the kink instability during the expansion phase
we compare  $t_{\rm kink}$ (Eq. \ref{eq:t_kink}) with the time that passed since 
the injection, $t_{\rm dyn}=z/c$, resulting in:
\begin{equation}\label{eq:t_kink_NR}
  \left.\frac{t_{\rm kink}}{t_{\rm dyn}}\right|_{{z}<{z}_{\rm max}}\simeq 200 f 
  \fracb{z}{z_1}^{\frac{3\xi-10}{10}}\fracb{\zh}{z_1}^{\frac{5-\xi}{10}}\fracb{\rL}{z_1}^{1/2}
  \fracb{2.5}{5-\xi}^{1/2}\left[1-\fracb{z}{\zh}^{\frac{5-\xi}{5}}\right]^{1/2}.
\end{equation}
For $z<z_{\rm max}$ this ratio scales like 
$z^{-0.25+0.3(\xi-2.5)}$,
therefore  it is minimal, or least stable, at  $z=z_{\rm max}$: 
\begin{equation}
  \left.\frac{t_{\rm kink}}{t_{\rm dyn}}\right|_{{z}={z}_{\rm max}}\simeq 10 f 
  \fracb{z_1}{2.5\times10^9}^{-1/2}r_{_{L,7}}^{1/2}\fracb{\zh}{z_1}^{\frac{\xi-2.5}{5}}
  \fracb{2.5}{5-\xi}^{1/2}\fracb{3\xi}{0.75(5+2\xi)}^{\frac{3\xi-10}{2(5-\xi)}}.
\end{equation}
Implying that the instability does't have enough time
to grow in the expanding part of the jet.
 
The fluid element stops expanding at  $z=z_{\rm max}$ and from there on  
its  radius  decreases until it reaches the head. The dynamical time 
available
for the instability to evolve now is $t_{\rm dyn}=\bar{z}/c$, resulting in the
ratio
\begin{equation}\label{eq:t_kink_NR}
  \left.\frac{t_{\rm kink}}{t_{\rm dyn}}\right|_{{z}>{z}_{\rm max}}\simgreat 200 f 
  \fracb{z_{\rm 
  max}}{z_1}^{3\xi/10}\fracb{\rL}{\bar{z}}^{1/2}\fracb{2.5}{5-\xi}^{1/2},
\end{equation}
where we approximated in $\rj$, $\bh(z)\simeq\bh(z_{\rm max})$, and 
$\zh/\bh - z/\bh(z)\simgreat \bar{z}$, which gives a lower limit to $t_{\rm 
kink}$. Using our fiducial parameters we get that 
$t_{\rm kink}/t_{\rm dyn}\sim 10 f (z_{\rm max}/\bar{z})^{1/2}$, which implies that 
the  time available for the instability to grow decreases  faster than the
growth time of the instability. Implying that the kink instability cannot 
grow in the  region where the jet's head 
is non-relativistic.

When the head is relativistic ($\uh\gg1$), the flow velocity is very close to the velocity of 
the head.
Here we cannot neglect the motion of the head during the propagation
of the fluid element. Namely, as a fluid element propagates from 
altitude $z$ to $z+\delta z$ the jet's head has moved from $z_h$
to $z_h+\beta_h\delta z$ which implies that the geometry of the
jet is changing during the motion of the fluid element.
Thus in this limit we need to use the total  derivative with respect to $z$:
\begin{equation}\label{eq:drjdz_rel}
  \frac{dr_j(z,\zh)}{dz} = \frac{\rj}{z} \left[\frac{\xi}{6} -\frac{1-\beta_h(\zh)}{4} 
  \frac{z}{\bar{z}}\right],
\end{equation}
where $\rj(z,\zh)$ is taken from Eq. (\ref{eq:r_j_NR}) and we approximated 
$d\rhead/dz\simeq\bh(z)$. The height of the maximal radius of a fluid element
is obtained by taking $d\rj/dz=0$. Giving
\begin{equation}
  \bar{z}_{\rm max}\equiv\zh-z({d\rj}/{dz}=0)\simeq 0.6 \zh (1-\bh)\frac{2.5}{\xi}
\end{equation}
Note that unlike in the case of a non-relativistic head, $\bar{z}_{\rm max}$ here is located 
above the widest point of the jet, which is a direct consequence of the fact that
the  jet's pattern expands at a comparable rate to the propagation rate of
the fluid element.

In the region where the fluid element expands, $z<z_{\rm max}$, the dynamical time 
is $t_{\rm dyn}$=z/c, similar to the non relativistic head.
In this case the ratio of $t_{\rm kink}/t_{\rm dyn}$ gives
\begin{equation}\label{eq:t_kink_R}
  \left.\frac{t_{\rm kink}}{t_{\rm dyn}}\right|_{{z}<{z}_{\rm max}}\simgreat140 f 
  \frac{\rL}{z_1}\fracb{z}{z_1}^{\frac{\xi-3}{3}}\fracb{\bar{z}}{\rL}^{1/2}.
\end{equation}
From Eq. (\ref{eq:drjdz_rel}) we can see that as long as $z\ll \bar{z}/(1-\bh)\simeq 
2\gh^2\bar{z}$, the fluid element expands like $\rj(t)\sim z^{\xi/6}$ when it 
moves from $z$ to $z+\delta z$.  Therefore during the 
expansion phase $t_{\rm kink}/t_{\rm dyn}\propto z^{-0.17 (3-\xi)/0.5}$, 
implying that instability growth is inhibited by the
increasing in the cylindrical radius. 
The jet is least stable at $z=z_{\rm max}$, where we get
\begin{equation}
  \left.{\frac{t_{\rm kink}}{t_{\rm dyn}}}\right|_{\bar{z}=\bar{z}_{\rm max}}
  \simgreat 4 f r_{_{\rm L,7}}^{1/2}z_{_{\rm h,11}}^{(\xi-3)/6}
  \fracb{z_1}{2.5\times10^9}^{-\xi/6}\fracb{\xi}{2.5}^{1/2},
\end{equation}
which implies again that the jet is most likely stable below $z_{\rm max}$.

Above $z_{\rm max}$ the radius of the fluid element decreases with $z$.
Since the head is propagating relativisticaly, 
the time that is left for the instability to grow is 
$t_{\rm dyn}(\bar{z}_{\rm max})=\bar{z}_{\rm max}/(1-\bh)\simeq 0.6 \zh \frac{2.5}{\xi}$.
Thus the time to reach the head is comparable to the time it takes to reach $z_{\rm max}$. 
We can therefore neglect the change in $\bh$ during this remaining time,
and write a  general expression for $t_{\rm dyn}(\bar{z})=\bar{z}/(1-\bh)$.
Comparing this time with $t_{\rm kink}$ we get the ratio 
\begin{equation}\label{eq:t_kink_R}
  \frac{t_{\rm kink}}{t_{\rm dyn}}\simeq 70 f 
  \fracb{\rL}{\bar{z}}^{1/2}\simeq 6 f r_{_{\rm L,7}}^{1/2}z_{_{\rm h,11}}^{(\xi-3)/6}
 \fracb{z_1}{2.5\times10^9}^{-\xi/6}\fracb{\xi}{2.5}^{1/2}\fracb{\bar{z}_{\rm max}}{\bar{z}}^{1/2},
\end{equation}
which scales as $\bar{z}^{-1/2}$, like in the case of a non-relativstic head.
We therefore conclude that unless $f$ is very small, the jet will not be disrupted by
kink instability as it propagates in the star. Even for 
values of $f\sim0.1$, the jet will only be marginally unstable and
is likely to survive the crossing of the star.

This analysis was preformed under the assumption that the jet is 
not rotating and moves upward rigidly.
In this case the poloidal field scales roughly like $r^{-1}$.
In a more realistic scenario of a rotating, differentially moving 
jet, the electric field plays 
a role in the transverse force balance. Then 
the outward gradient of the poloidal magnetic field could be
smaller than in the rigidly moving jet or even vanish altogether.
\citet{1996MNRAS.281....1I}
have shown that Poynting dominated jets with homogeneous
poloidal field are stable. \citet{1999MNRAS.308.1006L} 
considered the general case and found that
the instability growth rate decreases at shallower profiles 
of the poloidal magnetic
field and it goes to zero for a homogeneous field. Simulations of cylindrical jets with shallow
transverse distribution of the poloidal field \citep{2012ApJ...757...16M}
indeed revealed slower
perturbation growth in accord with linear
stability analysis. More importantly, they clearly demonstrate a nonlinear saturation
of the instability in this case so that the
initial cylindrical structure is not disrupted.
An important point is that simulations of jet launching by a spinning accreting black
hole reveal that in these Poynting-dominated jets, the poloidal
field is very close to uniform \citep{2008MNRAS.388..551T}.
It suggests that such Poynting dominated jets
could not be destroyed by the kink instability.
This agrees with 3D simulations by \citet{2009MNRAS.394L.126M}
who did not observe the kink instability in Poynting dominated jets
emanated from an accreting black hole. 

These findings add credibility to our conclusion that the jet is
stable. Nevertheless, we stress that our stability analysis is based on the results 
that are available in the literature, which use a somewhat different setup.
In particular in our model the jet is bounded by a tangential discontinuity
that separates it from the cocoon. The  discontinuity as well as the cocoon don't
 exist in the cited works.
This may affect both the growth rate and the non linear behavior of the kink 
instability, as well as potentially lead to additional instabilities.
All these effects can be fully tested with numerical 3D simulations.

\section{Explicit time dependence and the jet breakout time}\label{sec:t_b}

{The properties of the jet and the cocoon in our model depend
on two parameters, $t$ and $z$. So far it was convenient to include
the time dependence implicitly thorough the (time dependent) location
of the jet's head, $\zh(t)$. In this case we could also use the
coordinate $\bar z\equiv \zh(t)-z$ instead of $z$ in some situations.}
The explicit time dependence  can be obtained by calculating $t(\zh)$,
the time in which the jet's head reaches an altitude $\zh$:
\begin{equation} \label{eq:t_h}
  t(\zh) =
  \int_{0}^{\zh}\frac{dz}{\bh(z)c}\simeq\int_{0}^{z_1}\fracb{z_1}{z}^{\xi/5}\frac{dz}{c}+
  \int_{z_1}^{\zh}\left(1+\frac{1}{2}\fracb{z_1}{z}^{\xi/3}\right)\frac{dz}{c}.
\end{equation}
Here we care about  small differences from the speed of light
thus we approximate 
\begin{equation}\label{eq:beta_h_app}
  \frac{1}{\bh(z)} \simeq\left\{\matrix{
   a(z)^{-1/5}=(z_1/z)^{\xi/5} & \quad (z<z_1)\ \cr\cr
  \left[1-a(z)^{-1/3}\right]^{-1/2}\simeq1+1/2(z_1/z)^{\xi/3} &
\quad (z_1<z)\ .
  }\right.
\end{equation}
This gives the expression:
\begin{equation}\label{eq:t_h1}
  \frac{ct(\zh)}{\zh} \simeq\left\{\matrix{
   \frac{5}{5-\xi}\fracb{z_1}{\zh}^{\xi/5} & \quad (\zh<z_1)\ \cr\cr
  1+\frac{3}{2(3-\xi)}\fracb{z_1}{\zh}^{\xi/3}+
  \frac{z_1}{\zh}\left[\frac{\xi}{5-\xi}-\frac{3}{2(3-\xi)}\right] &
\quad (z_1<\zh)\ .
  }\right.
\end{equation}
{In the case where $z_h\gg z_1$ and $\xi<3$ the third term in Eq.
(\ref{eq:t_h1},b) is negligible, while for $\xi>3$ the second term
can be neglected, allowing for an explicit solution of $\zh(t)$.
Solving for $\zh$, and noting that when $\zh>z_1$,
$\gh^2(\zh)\simeq\fracb{\zh}{z_1}^{\xi/3}$ we get}
\begin{equation}\label{eq:zh_t}
z_h(t)\simeq\left\{\matrix{
z_1\left(\frac{5-\xi}{5}\frac{ct}{z_1}\right)^{5/(5-\xi)} & \quad (\zh<z_1)\ \cr\cr
\left[ct-z_1\left(\frac{\xi}{5-\xi}-\frac{3}{2(3-\xi)}\right)\right]
  \left[1+\frac{3}{2(3-\xi)\gh^2(z_h)}\right]^{-1}&
\quad (\zh>z_1)\ ,
}\right.
\end{equation}
This result can be substituted in Eqs. (\ref{eq:r_c} -- \ref{eq:r_j_NR}) in order 
to obtain the explicit time dependence of
all the model parameters.

A typical collapsar jet becomes relativistic at
\begin{equation}\label{eq:z_1}
  z_1\simeq a(R)^{-1/\xi}R = 2.5\times10^9 {\rm cm}
  \left[\fracb{3-\xi}{0.5}L_{50}^{-1}\fracb{M}{15M_\odot}r_{_{L,7}}^2\right]^{1/\xi}R_{11}^{1-3/\xi},
\end{equation}
Since $z_1\ll R$
the jet becomes relativistic deep in the star and therefore we can
approximate the time it breaks out of the stellar surface as
\begin{equation}\label{eq:t_R}
  t_R\equiv t_h(\zh=R)\simeq
  \frac{R}{c}\left(1+\frac{3}{2(3-\xi)a^{1/3}(R)}\right) =
 3.3R_{11}\left({1+0.14\fracb{0.5}{3-\xi}a_4^{-1/3}(R)}\right)~{\rm s}\ .
  \end{equation}
where
\eqb
a_4(R) \equiv \frac{a(R)}{10^4} = \fracb{0.5}{3-\xi}
L_{50}r_{L,7}^{-2}\fracb{M}{15\,M_\odot}^{-1}R_{11}^3\ .
\eqe
Thus the jet
reaches the stellar surface after $t_R\simeq 4~{\rm s}$, which is only
slightly longer than the light crossing time of the star.  

Last, we calculate the minimal activity time of the engine
that is required to push the jet out of the star.
When the jet propagates in the star, part of its energy in deposited
into the cocoon. It follows that the
engine needs to invest some minimal amount of energy to push the jet
out of the star, corresponding with a minimal activity time.  
The rest of the energy 
is available to produced the observed GRB emission. 
When the jet engine stops the information 
{that this has happened}
propagates upward at 
the fast
magnetosonic speed, in the fluid rest frame, which is 
very close to the speed of light. During that time the head
of the jet continues to move forward as if nothing has happened.
If this information reaches the head before the head breaks through the
stellar surface, the jet will slow down and fail to breakout.
We define the threshold activity time for a successful jet breakout
, $t_{\rm th} = t_R-R/c$, as the
difference between the breakout time, $t_R$, and the light crossing time,
$R/c$, of the star.
{If the engine stops at $t<t_{\rm th}$ the jet will fail to exit the star.}
In cases when the head of the jet is
non-relativistic during the entire crossing, $t_R\gg R/c$; then
$t_{th}\simeq t_R$.  When the jet's head is relativistic the
threshold time can be easily obtained by noting that $ct_{\rm th}/R = (ct_R/R)-1$
where $ct_R/R$ is given by  Eq. (\ref{eq:t_R}):
\begin{equation}
  t_{\rm th}\simeq\frac{2}{2(3-\xi)}\frac{R}{c}a(R)^{-1/3}\simeq
  0.5\left[\fracb{0.5}{3-\xi}^{2}L_{50}^{-1}\fracb{M}{15M_\odot}r_{_{L,7}}^{2}\right]^{1/3}
  ~ {\rm s}.
\end{equation}
{As the jet becomes relativistic deep in
the star this time is much shorter than $t_R$, the time it takes for  the jet's head
to reaches the stellar surface. }
Note that $t_{th}$ is the last time where
any information that leaves the engine can reach the head
before the jet breaks out of the star. Therefore we can estimate the maximal amount
of energy that the jet can put into the stellar envelope as:
\begin{equation}
  E_{c,\rm max}=2\Lj t_{\rm th}\simeq10^{50}~{\rm erg}\times
  \left[\fracb{0.5}{3-\xi}^{2}L_{50}^{-1}\fracb{M}{15M_\odot}r_{_{L,7}}^{2}\right]^{1/3},
\end{equation} 
where the factor of 2 is used to account for the two sides of the jet.
This energy is well below the energy
required to unbind the star, which is of the order of $\sim
1.5\times10^{51}$ ergs for our fiducial set of parameters. 
Implying that magnetic GRB jets by themselves cannot unbind 
their progenitor stars.

\section{Conclusions}\label{sec:conclusions}

We have developed an analytic model for the propagation of a relativistic Poynting flux
dominated 
jet in a dense medium and applied it to
collapsar jets. {Namely, }
to relativistic  jets produced within a collapsing star that   propagate in their massive star
progenitor. {Note that we consider as collapsars all cases of 
GRB models  that involves jets that propagate
though their massive stellar progenitor, regardless of the nature of the central engine that 
powers them.}

As the jet propagates it pushes
the material in front of it, leading to the formation of a bow shock
ahead of the jet. Matter that passes through this shock is heated and
and forms a cocoon
,that applies pressure on the jet and collimates it.
At the same time the cocoon propagates
sidewards into the stellar envelope.
Inside the star the cocoon pressure is large enough to
collimate the jet close to the launching point so that the
 jet's outer boundary maintains a strong lateral causal contact with
its center. In this case the jet is collimated smoothly without
dissipation, and therefore almost all of the magnetic flux reaches the
head.

We have shown that both the  size, $\rhead$, and the proper velocity, $\uh$   of the jet's head depend only on the ratio $(L_j/r_L^2) (1/\rho_{ext})$.
The first factor depends on the total luminosity  of the jet and the size of its central engine  while the latter depends only on the  external density profile.  These quantities do not depend on the details of the  jet structure
behind its head.   Since the energy is injected into the cocoon only at the
jet's head, this enables us to analytically estimate the energy
injection rate into the cocoon without the need to know the  exact structure of the jet.
Using this energy injection rate we  calculate  the cocoon's structure. Finally, once the 
cocoon pressure profile is known
we determine the jet's structure. We show that the resulting jet structure is indeed consistent with the
properties of the jet's head.
In spite of the various approximations used,
the analytic model is in a good agreement with the results  
 (see Figure \ref{fig:jet_simulation}) of recent  simulations \citep{BT14}.

The propagation of the Poynting flux dominated jet described here should be contrasted with 
the propagation of a similar hydrodynamic jet.
The basic difference arises due to the fact that while strong shocks can form in a 
hydrodynamic flow they cannot arise in
a Poynting flux dominated outflow 
\citep{1984ApJ...283..710K}.
Therefore in a hydrodynamic 
jet
the head of the jet involves both a forward shock that propagates into the surrounding stellar 
material and a reverse shock that propagates into the jet's material and slows it down 
significantly. At the same time the pressure exerted by the cocoon on the jet leads to a 
collimation shock that takes place deep near the base of the jet 
\citep{1997MNRAS.288..833K,2007ApJ...671..678B,2011ApJ...740..100B}.
This shock heats initially cold matter and the slows it down. The interplay between these 
two shocks, the one at the head of the jet and the collimation shock at the base determines 
the hydrodynamic jet structure. Unlike the Poynting flux dominated jet the hydrodynamic jet 
is slow and for most relevant parameters its head propagates with sub-relativistic velocities.
This means that the jet crossing time is longer and the energy deposited by the shock and 
given to the stellar envelope is larger by an order of magnitude than the corresponding 
quantities for the corresponding Poynting flux dominated outflow.


The most questionable assumption is that the magnetic jet remains 
axi-symemtric and stable.
Our estimates have shown that the kink instability, which is the fastest growing 
instability,  is unlikely to distrupt the jet while it propagates in the star.
It could, however,  somewhat change the details of our results, for
example  it may increase somewhat the size the head and therefore decreases its velocity and
subsequently increases the jet escape time.
However our analysis is based on linear estimates of  the growth rate of
the instability and even those are somewhat approximate.
Clearly only full three dimensional simulations could clarify the issue of stability.


\citet{2013ApJ...764..148L}  have analyzed recently a similar system reaching a 
very different conclusions. They
have argued that such a
magnetic dominated jet becomes unstable while propagating within the star. 
Due to this instability
the jet dissipates its magnetic energy on a time scale that is much shorter than 
the jet propagation time. The  jet continues propagating as a hydrodynamic
jet within the star. 
The main difference between their analysis and ours  is that 
\citet{2013ApJ...764..148L}
assume  that the inner part of the cocoon is  dominated by the pressure of
 toroidal magnetic fields, exerting a much larger pressure on the
jet boundary. In addition they assume that the cocoon pressure
is uniform in the z direction. These two assumptions result in a cylindrical
jet having a width that is comparable to $\rhead$. Such a narrow jet is  
unstable to kink instability. Overall we expect that the assumption of a 
toroidal magnetic pressure dominated region in the cocoon is not entirely justified. 
In particular numerical simulations \citep{BT14} do not show such a structure.

When applying these results to  collapsars we recall that 
the typical size of the head is  a few light cylinder radii
$\rL$  and the  propagation proper velocity is  mildly relativistic
($\uh\sim\rhead/\rL\sim$ a few)  during most of the propagation inside
the star. Thus, the jet crosses the star in a time that is close to its
light crossing time, $R/c$. The fast head velocity implies that once
the jet is launched it will most likely exit the star. The fast crossing time also implies that the
energy injected into the star is minimal. Astrophysical implications of these finds
will be discussed elsewhere \citep{BGPL14}.

\subsection*{Acknowledgements}
We thank  A. Tchekhoskoy, M. Kuntz, A Spitkowsky,
J. Stone and A. Levinson for helpful discussions and the  ISSI (Bern) for hospitality while the final version of
this paper was written.
This research was supported by the ERC advanced research grant ``GRBs''
by the  I-CORE (grant No 1829/12), by HUJ-USP grant and by
the Max-Planck/Princeton Center for Plasma Physics. 

\appendix
\section{Basic derivation of the dependence of the jet properties on the inlet
velocity of the flow}\label{app:basic eq}

Let us consider the flow between two flux surfaces corresponding to
$r(z)$ and $r(z)+\delta r(z)$ (see figure \ref{fig:App}). The stream lines are
along the poloidal field lines, in a direction $\hat{e}_p$ that makes an angle
$\alpha$ with the $z$-axis (see Fig.~\ref{fig:App}) so that
\begin{equation}\label{eq:alpha}
\cos\alpha = 
\left[1+\left(\frac{dr}{dz}\right)^2\right]^{-1/2}
\ ,\quad\quad
\sin\alpha = 
\left[1+\left(\frac{dz}{dr}\right)^2\right]^{-1/2}\ .
\end{equation}

Let us now consider a fluid element of height $\delta l$ along the
poloidal direction. Flux freezing implies that the magnetic
flux through a surface normal to the relevant magnetic field component
is constant, implying that
${\bf B_\phi}\cdot({\bf \delta r}\times{\bf \delta l})=B_\phi\delta r\delta
l\cos\alpha={\rm const}$ ; ${\bf B_p}\cdot (r\delta r){\bf\hat z}=B_pr\delta
r\cos\alpha={\rm const}$, and therefore
$B_\phi/B_p \propto r/\delta l$, where $B_\phi$ and $B_p$ are the
azimuthal and poloidal magnetic field components, respectively.  The
mass of the fluid element scales as $\rho\Gamma\delta l r\delta
r\cos\alpha$, while the rest mass flux (which is also conserved
between two stream surfaces in steady state) scales as $\rho u r\delta
r\cos\alpha$ (where $u =\Gamma\beta$ is the proper velocity). By taking
the ratio of these two conserved quantities one obtains $\delta l
\propto\beta$, which implies that $B_\phi/B_p
\propto r/\beta$.

\begin{figure}
  \centering
  \includegraphics[width=5cm]{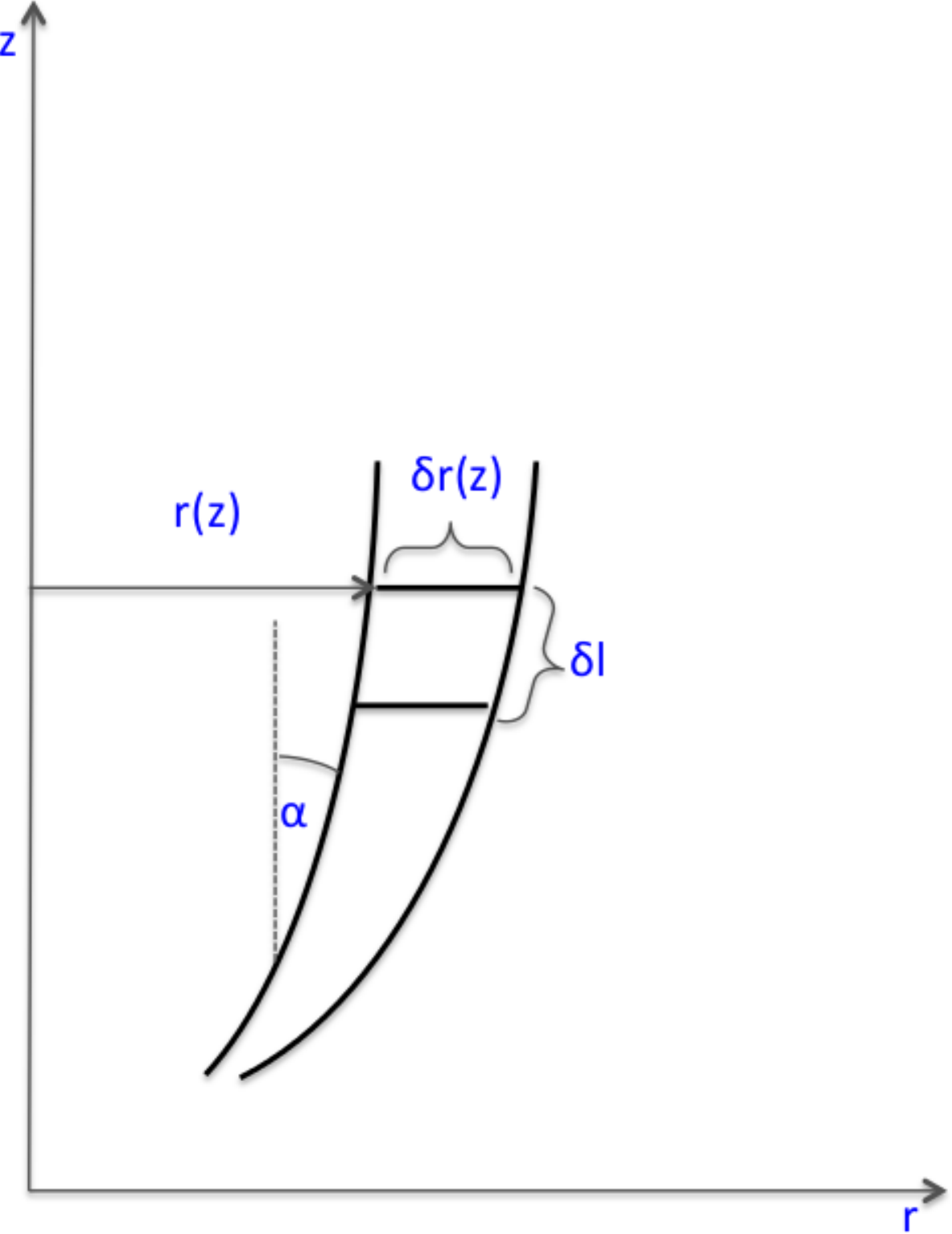}
  \caption{A schematic drawing of the jet element}
  \label{fig:App}
\end{figure}

Now, since the jet is in lateral causal contact and thus in lateral
equilibrium, the two components of the field are comparable in the
comoving fame, $B'_\phi \sim B'_p$. Lorentz transformation of the
fields to the lab frame (in which the central source and external
medium are at rest), implies $B_\phi/\Gamma = B'_\phi\sim B'_p = B_p$,
so that $\Gamma\sim B_\phi/B_p \propto r/\beta$ and $r \propto u =
\Gamma\beta$.  Near the light cylinder, at $r\sim \rL$, the two field
components are comparable in the lab frame and $u(\rL) \sim 1$, which
implies that $u \sim r/\rL$.  In the limit where $\cos\alpha\simeq1$
we get that the magnetic pressure $p_{_B}\sim (B'_\phi)^2\sim
(B_\phi/\Gamma)^2 \sim (\delta r u)^{-2}\sim r^{-4}$.

When the head of the jet is Newtonian ($\uh < 1$), the velocity of the 
jet material is non-relativistic in the part of the jet that holds 
 $\rj / \rL \simeq \uj < 1$. In this region the approximation of a steady state in 
the lab frame breaks down. However, since most of
the work done by the jet on the cocoon is performed at its head where, 
$\rj \leq \rhead \simeq \rL\uh$,
then in the region $\rhead < \rj \leq \rL$ the energy flux in the lab frame is still constant
and equal to $\Lj$ to a good approximation. Therefore, one can use Eq. (\ref{eq:Lj}) 
which shows that $\Lj\propto \rj^2\Gj\uj\pj\propto \pj\rj^4/\bj$. Implying that
$\pj\propto\bj\rj^{-4}$.
 In the relativistic section of the
jet ($\rj>\rL$), where $\bj\approx1$, this recovers the above result of $\pj\propto\rj^{-4}$ 
(or $p_b\propto r^{-4}$, which shows that there the steady state approximation 
in the lab frame works well).
In the Newtonian part of the jet, however ($\rj<\rL$), we get that $\bj\approx\uj\propto\rj$, 
therefore the scaling of
the pressure with the jet radius changes to $\pj\propto\rj^{-3}$ (or $p_b\propto r^{-3}$).



\begin{thebibliography}{50}
\expandafter\ifx\csname natexlab\endcsname\relax\def\natexlab#1{#1}\fi

\bibitem[{{Appl}, {Lery} \& {Baty}(2000){Appl}, {Lery}, \&
  {Baty}}]{2000A&A...355..818A}
{Appl} S., {Lery} T., {Baty} H., 2000, \aap, 355, 818

\bibitem[{{Begelman}(1998)}]{1998ApJ...493..291B}
{Begelman} M.~C., 1998, \apj, 493, 291

\bibitem[{{Beloborodov} \& {Uhm}(2006)}]{2006ApJ...651L...1B}
{Beloborodov} A.~M., {Uhm} Z.~L., 2006, \apjl, 651, L1

\bibitem[{{Beskin}, {Kuznetsova} \& {Rafikov}(1998){Beskin}, {Kuznetsova}, \&
  {Rafikov}}]{1998MNRAS.299..341B}
{Beskin} V.~S., {Kuznetsova} I.~V., {Rafikov} R.~R., 1998, \mnras, 299, 341

\bibitem[{{Bromberg} {et~al}\mbox{.}(2014){Bromberg}, {Granot}, {Piran}, \&
  {Lyubarsky}}]{BGPL14}
{Bromberg} O., {Granot} J., {Piran} T., {Lyubarsky} Y., 2014, in prep

\bibitem[{{Bromberg} \& {Levinson}(2007)}]{2007ApJ...671..678B}
{Bromberg} O., {Levinson} A., 2007, \apj, 671, 678

\bibitem[{{Bromberg} {et~al}\mbox{.}(2011){Bromberg}, {Nakar}, {Piran}, \&
  {Sari}}]{2011ApJ...740..100B}
{Bromberg} O., {Nakar} E., {Piran} T., {Sari} R., 2011, \apj, 740, 100

\bibitem[{{Bromberg} {et~al}\mbox{.}(2012){Bromberg}, {Nakar}, {Piran}, \&
  {Sari}}]{2012ApJ...749..110B}
{Bromberg} O., {Nakar} E., {Piran} T., {Sari} R., 2012, \apj, 749, 110

\bibitem[{{Bromberg} \& {Tchekhovskoy}(2014)}]{BT14}
{Bromberg} O., {Tchekhovskoy} A., 2014, in prep

\bibitem[{{Eichler}(1993)}]{1993ApJ...419..111E}
{Eichler} D., 1993, \apj, 419, 111

\bibitem[{{Giannios} \& {Spruit}(2006)}]{2006A&A...450..887G}
{Giannios} D., {Spruit} H.~C., 2006, \aap, 450, 887

\bibitem[{{Granot}, {Cohen-Tanugi} \& {do Couto e Silva}(2008){Granot},
  {Cohen-Tanugi}, \& {do Couto e Silva}}]{2008ApJ...677...92G}
{Granot} J., {Cohen-Tanugi} J., {do Couto e Silva} E., 2008, \apj, 677, 92

\bibitem[{{Granot}, {Komissarov} \& {Spitkovsky}(2011){Granot}, {Komissarov},
  \& {Spitkovsky}}]{2011MNRAS.411.1323G}
{Granot} J., {Komissarov} S.~S., {Spitkovsky} A., 2011, \mnras, 411, 1323

\bibitem[{{Istomin} \& {Pariev}(1996)}]{1996MNRAS.281....1I}
{Istomin} Y.~N., {Pariev} V.~I., 1996, \mnras, 281, 1

\bibitem[{{Kawanaka}, {Piran} \& {Krolik}(2013){Kawanaka}, {Piran}, \&
  {Krolik}}]{2013ApJ...766...31K}
{Kawanaka} N., {Piran} T., {Krolik} J.~H., 2013, \apj, 766, 31

\bibitem[{{Kennel} \& {Coroniti}(1984)}]{1984ApJ...283..710K}
{Kennel} C.~F., {Coroniti} F.~V., 1984, \apj, 283, 710

\bibitem[{{Kohler} \& {Begelman}(2012)}]{2012MNRAS.426..595K}
{Kohler} S., {Begelman} M.~C., 2012, \mnras, 426, 595

\bibitem[{{Komissarov} {et~al}\mbox{.}(2007){Komissarov}, {Barkov}, {Vlahakis},
  \& {K{\"o}nigl}}]{2007MNRAS.380...51K}
{Komissarov} S.~S., {Barkov} M.~V., {Vlahakis} N., {K{\"o}nigl} A., 2007,
  \mnras, 380, 51

\bibitem[{{Komissarov} \& {Falle}(1997)}]{1997MNRAS.288..833K}
{Komissarov} S.~S., {Falle} S.~A.~E.~G., 1997, \mnras, 288, 833

\bibitem[{{Komissarov} {et~al}\mbox{.}(2009){Komissarov}, {Vlahakis},
  {K{\"o}nigl}, \& {Barkov}}]{2009MNRAS.394.1182K}
{Komissarov} S.~S., {Vlahakis} N., {K{\"o}nigl} A., {Barkov} M.~V., 2009,
  \mnras, 394, 1182

\bibitem[{{Lazzati} \& {Begelman}(2005)}]{2005ApJ...629..903L}
{Lazzati} D., {Begelman} M.~C., 2005, \apj, 629, 903

\bibitem[{{Levinson} \& {Begelman}(2013)}]{2013ApJ...764..148L}
{Levinson} A., {Begelman} M.~C., 2013, \apj, 764, 148

\bibitem[{{Lithwick} \& {Sari}(2001)}]{2001ApJ...555..540L}
{Lithwick} Y., {Sari} R., 2001, \apj, 555, 540

\bibitem[{{Lyubarskii}(1999)}]{1999MNRAS.308.1006L}
{Lyubarskii} Y.~E., 1999, \mnras, 308, 1006

\bibitem[{{Lyubarskij}(1992)}]{1992SvAL...18..356L}
{Lyubarskij} Y.~E., 1992, Soviet Astronomy Letters, 18, 356

\bibitem[{{Lyubarsky}(2009)}]{2009ApJ...698.1570L}
{Lyubarsky} Y., 2009, \apj, 698, 1570

\bibitem[{{Lyubarsky}(2010)}]{2010ApJ...725L.234L}
{Lyubarsky} Y., 2010, \apjl, 725, L234

\bibitem[{{Lyubarsky}(2011)}]{2011PhRvE..83a6302L}
{Lyubarsky} Y., 2011, \pre, 83, 016302

\bibitem[{{MacFadyen} \& {Woosley}(1999)}]{1999ApJ...524..262M}
{MacFadyen} A.~I., {Woosley} S.~E., 1999, \apj, 524, 262

\bibitem[{{MacFadyen}, {Woosley} \& {Heger}(2001){MacFadyen}, {Woosley}, \&
  {Heger}}]{2001ApJ...550..410M}
{MacFadyen} A.~I., {Woosley} S.~E., {Heger} A., 2001, \apj, 550, 410

\bibitem[{{Matzner}(2003)}]{2003MNRAS.345..575M}
{Matzner} C.~D., 2003, \mnras, 345, 575

\bibitem[{{McKinney} \& {Blandford}(2009)}]{2009MNRAS.394L.126M}
{McKinney} J.~C., {Blandford} R.~D., 2009, \mnras, 394, L126

\bibitem[{{Michel}(1969)}]{1969ApJ...158..727M}
{Michel} F.~C., 1969, \apj, 158, 727

\bibitem[{{Mignone} {et~al}\mbox{.}(2010){Mignone}, {Rossi}, {Bodo}, {Ferrari},
  \& {Massaglia}}]{2010MNRAS.402....7M}
{Mignone} A., {Rossi} P., {Bodo} G., {Ferrari} A., {Massaglia} S., 2010,
  \mnras, 402, 7

\bibitem[{{Mizuno} {et~al}\mbox{.}(2009){Mizuno}, {Lyubarsky}, {Nishikawa}, \&
  {Hardee}}]{2009ApJ...700..684M}
{Mizuno} Y., {Lyubarsky} Y., {Nishikawa} K.-I., {Hardee} P.~E., 2009, \apj,
  700, 684

\bibitem[{{Mizuno} {et~al}\mbox{.}(2012){Mizuno}, {Lyubarsky}, {Nishikawa}, \&
  {Hardee}}]{2012ApJ...757...16M}
{Mizuno} Y., {Lyubarsky} Y., {Nishikawa} K.-I., {Hardee} P.~E., 2012, \apj,
  757, 16

\bibitem[{{Mizuta} \& {Aloy}(2009)}]{2009ApJ...699.1261M}
{Mizuta} A., {Aloy} M.~A., 2009, \apj, 699, 1261

\bibitem[{{Mizuta} \& {Ioka}(2013)}]{2013ApJ...777..162M}
{Mizuta} A., {Ioka} K., 2013, \apj, 777, 162

\bibitem[{{Morsony}, {Lazzati} \& {Begelman}(2007){Morsony}, {Lazzati}, \&
  {Begelman}}]{2007ApJ...665..569M}
{Morsony} B.~J., {Lazzati} D., {Begelman} M.~C., 2007, \apj, 665, 569

\bibitem[{{O'Neill}, {Beckwith} \& {Begelman}(2012){O'Neill}, {Beckwith}, \&
  {Begelman}}]{2012MNRAS.422.1436O}
{O'Neill} S.~M., {Beckwith} K., {Begelman} M.~C., 2012, \mnras, 422, 1436

\bibitem[{{Piran}(1995)}]{1995astro.ph..7114P}
{Piran} T., 1995, ArXiv Astrophysics e-prints

\bibitem[{{Porth}, {Komissarov} \& {Keppens}(2014){Porth}, {Komissarov}, \&
  {Keppens}}]{2014MNRAS.438..278P}
{Porth} O., {Komissarov} S.~S., {Keppens} R., 2014, \mnras, 438, 278

\bibitem[{{Spruit}, {Foglizzo} \& {Stehle}(1997){Spruit}, {Foglizzo}, \&
  {Stehle}}]{1997MNRAS.288..333S}
{Spruit} H.~C., {Foglizzo} T., {Stehle} R., 1997, \mnras, 288, 333

\bibitem[{{Tchekhovskoy}, {McKinney} \&
  {Narayan}(2008{\natexlab{a}}){Tchekhovskoy}, {McKinney}, \&
  {Narayan}}]{2008AIPC.1054...71T}
{Tchekhovskoy} A., {McKinney} J.~C., {Narayan} R., 2008{\natexlab{a}}, in
  American Institute of Physics Conference Series, Vol. 1054, American
  Institute of Physics Conference Series, {Axelsson} M., ed., pp. 71--77

\bibitem[{{Tchekhovskoy}, {McKinney} \&
  {Narayan}(2008{\natexlab{b}}){Tchekhovskoy}, {McKinney}, \&
  {Narayan}}]{2008MNRAS.388..551T}
{Tchekhovskoy} A., {McKinney} J.~C., {Narayan} R., 2008{\natexlab{b}}, \mnras,
  388, 551

\bibitem[{{Tchekhovskoy}, {McKinney} \& {Narayan}(2009){Tchekhovskoy},
  {McKinney}, \& {Narayan}}]{2009ApJ...699.1789T}
{Tchekhovskoy} A., {McKinney} J.~C., {Narayan} R., 2009, \apj, 699, 1789

\bibitem[{{Tchekhovskoy}, {Narayan} \& {McKinney}(2010){Tchekhovskoy},
  {Narayan}, \& {McKinney}}]{2010NewA...15..749T}
{Tchekhovskoy} A., {Narayan} R., {McKinney} J.~C., 2010, \na, 15, 749

\bibitem[{{Woosley}(1993)}]{1993ApJ...405..273W}
{Woosley} S.~E., 1993, \apj, 405, 273

\bibitem[{{Zakamska}, {Begelman} \& {Blandford}(2008){Zakamska}, {Begelman}, \&
  {Blandford}}]{2008ApJ...679..990Z}
{Zakamska} N.~L., {Begelman} M.~C., {Blandford} R.~D., 2008, \apj, 679, 990

\bibitem[{{Zhang}, {Woosley} \& {MacFadyen}(2003){Zhang}, {Woosley}, \&
  {MacFadyen}}]{2003ApJ...586..356Z}
{Zhang} W., {Woosley} S.~E., {MacFadyen} A.~I., 2003, \apj, 586, 356

\end{thebibliography}
\bibliographystyle{mn2e.bst} 



\end{document}